\documentclass[preprint,aps]{revtex4}
\usepackage[dvipdfmx]{graphicx}
\def\vc#1{\mbox{\boldmath $#1$}}
\usepackage{wick}
\usepackage[dvipdfmx]{color}
\usepackage{amsmath}

\begin{document}

\title{
Tensor optimized Fermi sphere method for nuclear matter\\ 
$-$ power series correlated wave function and a cluster expansion $-$}

\author{Taiichi~Yamada}\email[]{yamada@kanto-gakuin.ac.jp}

\affiliation{College of Science and Engineering, Kanto Gakuin University, Yokohama 236-8501, Japan} 

\begin{abstract}%
A new formalism, called ``tensor optimized Fermi sphere (TOFS) method'', is developed to treat the nuclear matter using a bare interaction among nucleons.
In this method, the correlated nuclear matter wave function is taken to be a power series type, $\Psi_{N}=[\sum_{n=0}^{N} {(1/n!)F^n}]\Phi_0$ and an exponential type, $\Psi_{\rm ex}=\exp(F) \Phi_0$, with the uncorrelated Fermi-gas wave function $\Phi_0$, where the correlation operator $F$ can induce central, spin-isospin, tensor, etc.~correlations, and $\Psi_{\rm ex}$ corresponds to a limiting case of $\Psi_{N}$ ($N \rightarrow \infty$). 
In the TOFS formalism based on Hermitian form, it is shown that the energy per particle in nuclear matter with $\Psi_{\rm ex}$ can be expressed in terms of a linked-cluster expansion.
On the basis of these results, we present the formula of the energy per particle in nuclear matter with $\Psi_{N}$.
We call the $N$th-order TOFS calculation for evaluating the energy with $\Psi_N$, where the correlation functions are optimally determined in the variation of the energy.
{The TOFS theory is applied for the study of symmetric nuclear matter using a central \textit{NN} potential with short-range repulsion.
The calculated results are fairly consistent to those of other theories such as the Brueckner-Hartree-Fock approach etc.
}
\end{abstract}

\maketitle

\section{Introduction}
\label{sec:Introduction}

One of the fundamental problems in nuclear physics is to understand the properties of homogeneous nuclear and neutron matter on the basis of the bare interaction among nucleons.
Their information at high density plays a crucial role in the determination of the structure of neutron-star interiors~\cite{glendenning,raffelt96}.
On the other hand, $\alpha$-particle (quartet) condensation, induced by the strong quartet correlations, is predicted to occur at lower density region of symmetric nuclear matter~\cite{roepke98,beyer00,takemoto04,sogo09,sogo10-1,sogo10-2}.
This is related to $\alpha$-cluster states in finite system, such as the Hoyle state in $^{12}$C, observed in the excited energy region of light nuclei~\cite{wildermuth77,ikeda80,tohsaki00,yamada11,horiuchi12}.

Over the year several many-body theories have been developed to describe symmetric nuclear and neutron matter, since Brueckner et al.~presented their famous theories in 1950s, ie.\ the Brueckner theory~\cite{brueckner55_1,brueckner55_2,brueckner58} and Brueckner-Bethe-Goldstone theory~\cite{bethe56,bethe57,goldstone57}.
In the non-relativistic framework, the following typical approaches have been proposed so far:~The Brueckner-Hartree-Fock (BHF) approach~\cite{brueckner58,mahayux,baldo91,schulze95}, the Brueckner-Bethe-Goldstone (BBG) approach up to the third order in hole-line expansion~\cite{day67,day83,song98,baldo00,baldo01,sartor06}, the self-consistent Green's function (SCGF) method~\cite{dickhoff04,frick05,soma06,rios09},  the auxiliary field diffusion Monte Carlo (AFDMC)~\cite{gandolfi07,gandolfi09}, the Green's function Monte Carlo (GFMC)~\cite{carlson03}, the Fermi hypernetted chain (FHNC) method~\cite{iwamoto57,fantoni72,fantoni74,fantoni78,pandharipande79,akmal98}, the coupled-cluster (CC) method~\cite{baardsen13,hagen14}, the quantum Monte Carlo lattice calculation~\cite{abe09}, and so on.
Using the FHNC method with the modern nuclear Hamiltonian composed of the Argonne V18 (AV18) two-nucleon interaction~\cite{wiringa95} and the Urbana IX three-nucleon interaction~\cite{pudliner95}, Akmal, Pandharipande, and Ravenhall evaluated the ground-state energies for symmetric nuclear matter and neutron matter, taking into account the boost effect caused by the relativistic kinematics~\cite{akmal98}.
On the other hand, in the relativistic framework, 
the Dirac-Brueckner-Hartree-Fock  (DBHF) method has been proposed by Brockmann and Machleidt~\cite{brockmann90}, where the meson exchange interaction (Bonn potential) was used for the two nucleon interaction.

It is interesting to compare the results of the energy per particle for nuclear and neutron matter $E(\rho)/A$ as a function of matter density $\rho$, calculated in several many-body approaches proposed so far with a realistic two-nucleon interaction.
Regarding this issue, the comparative study with the Argonne two-nucleon potentials (AV4', AV6', AV8', and AV18)~\cite{wiringa95} has been performed~\cite{baldo12} for the BHF approach, BBG approach, SCGF method, AFDMC, GFMC, and FHNC, where the three-nucleon interaction is not included.
It is noted that AV4' is the central-force type, AV6' consists of only the central and tensor forces, AV8' does of the central, tensor, and spin-orbit forces, and AV18 is one of the modern two-nucleon interaction expressed as a sum of 18 operators, which provides an excellent fit to all of the nucleon-nucleon scattering data in the Nijmegen database. 
The results of $E(\rho)/A$ are shown in Figs.~3 and 4 of Ref.~\cite{baldo12}.
In the neutron matter, the difference of the behavior of $E(\rho)/A$ among various  methods (BHF, SCGF, FHNC, AFDMC, BBG, GFMC) is likely to be small for any Argonne potential, although the discrepancies among them are gradually enhanced at higher density.
In the symmetric nuclear matter, however, one can see non-negligible dependence of the behavior of $E(\rho)/A$ on various methods.
For example, in the cases of AV8' and AV6', the significant dependence of the behavior of $E(\rho)/A$ on the calculated methods (BHF, SCGF, FHNC, AFDMC, BBG) exist even in the lower density region including the normal density $\rho_0$, and they are amplified in higher density, although the dependence in the case of the AV4' potential (central force) is quite small. 
The contribution from the non-central forces, in particular, tensor force (and spin-orbit one), is more important in the symmetric nuclear matter than the neutron matter.
Thus the dependence of $E(\rho)/A$ on the calculated methods is likely to be caused mainly by the different treatment of the non-central components in medium, in particular, tensor component (and spin-orbit one) in each methods, together with the treatment of many-body terms appearing in the nuclear matter calculations. 
The above mentioned facts suggest the matter of convergence of $E(\rho)/A$.

Recently, a new variational framework for ab initio description of light nuclei, called ``tensor-optimized antisymmetrized molecular dynamics (TOAMD)'', has been proposed~\cite{myo15,myo17_1,myo17_2,myo17_3,myo17_4,myo17_5}.
The AMD wave function~\cite{enyo03,enyo12} is used as the uncorrelated wave function, and the correlation functions for the central-operator and tensor-operator types, $F_S$ and $F_D$, respectively, are introduced and employed in power series form of the wave function, which is different from the Jastrow method.
In the analysis of {\it s}-shell nuclei with TOAMD, they found that the results of Green's function Monte Carlo (GFMC) are nicely reproduced within the double products of the correlation functions, where each correlation function in every term is independently optimized in the variation of total energy.
Quite recently, other methods named as ``high-momentum AMD (HM-AMD)''  as well as the hybridization of TOAMD and HM-AMD called ``HM-TOAMD'' have proposed to study the structure of light nuclei~\cite{myo17_6,lyu17}.
 
Here, it is an intriguing issue to explore a framework for describing nuclear matter with the power series  correlated wave function.
In the FHNC framework, the nuclear matter is described by the Jastrow-type correlated wave  function, and the energy per particle is evaluated in a diagrammatic cluster expansion.
Fantoni and Rosani proved a linked-cluster expansion theorem that in the Jastrow-type nuclear matter wave function only the linked diagrams contributes to the energy and the unlinked ones are canceled out at each order of the cluster expansion~\cite{fantoni72,fantoni74,fantoni78,pandharipande79}.
In the TOAMD framework describing finite nuclei, one can take an arbitrary power series form with respect to the correlation functions, $F_S$ and $F_D$, which are variationally determined in the energy.
However, it is non-trivial that any power series form is allowed for describing the nuclear matter from the light of the cluster expansion for the energy. 
One needs to seek a formalism satisfying a sort of the linked-cluster expansion theorem in the case of using the power series correlated wave function.


The main purpose of the present article is to provide a new formalism to treat the nuclear matter within the framework of a power series correlated wave function $\Psi_{N}$ together with an exponential type $\Psi_{\rm ex}$:~$\Psi_{N} = [\sum_{n=0}^{N} (1/n!) F^n] \Phi_0$ and $\Psi_{\rm ex} = \exp(F) \Phi_0$, with $F = F_{S}+F_{D}$, where the latter is treated as a limiting case ($N \rightarrow \infty$) for the former, and $\Phi_0$ is the uncorrelated Fermi-gas wave function.
In the present framework, the tensor correlation $F_D$ is treated the same as the central one $F_S$, since both the correlations play an important role in nuclear matter as well as finite nuclei.
This new method based on Hermitian form is called ``tensor optimized Fermi sphere (TOFS)''.
It is proved that the energy per particle in nuclear matter with $\Psi_{\rm ex}$ is expressed as the sum of the linked diagrams with use of the cluster expansion in terms of $F$, and the unlinked diagrams are canceled out at each order of the cluster expansion.
In other words, a sort of the linked-cluster expansion theorem is established in the TOFS framework with use of $\Psi_{\rm ex}$. 
Based on these results, the formula of the energy per particle in nuclear matter with $\Psi_{N}$ is given as an approximation of that with $\Psi_{\rm ex}$.  
This formula  is useful to perform actual numerical calculations of nuclear matter. 
Evaluating the energy per particle in nuclear matter with $\Psi_N$ is called the $N$th-order TOFS calculation, where it converges to that with $\Psi_{\rm ex}$ in limiting case of $N \rightarrow \infty$.
In the TOFS calculation, the correlation functions, $F_S$ and $F_D$, are expanded into the Gaussian functions.
The expansion coefficients and Gaussian ranges are treated as the variational parameters for evaluating the energy per particle in nuclear matter, where we take into account of all the matrix elements of the many-body operators coming from the product of operators, $F^{n_1}\mathcal{H}F^{n_2}$, with $0 \le n_1,n_2 \le N$, where $\mathcal{H}$ is the Hamiltonian of nuclear matter.
The present TOFS framework is in contrast to the framework of the TOAMD and HM-AMD, where the correlated nuclear wave function is described by the arbitrary power series function with respect to $F_S$ and $F_D$ multiplied to the uncorrelated AMD wave function, as mentioned above. 
In nuclear matter the arbitrary power series function with respect to $F_S$ and $F_D$ is not allowed for the correlated nuclear matter wave function, because in this case the unlinked diagrams are not canceled out completely, as shown in the present study.

The present paper is organized as follows:\ In Sec.~\ref{sec:2}, we formulate in detail the power series correlated wave function of nuclear matter $\Psi_{N}$ together with the exponential one $\Psi_{\rm ex}$.
A cluster expansion and linked diagrams in the TOFS framework are discussed in Sec.~\ref{sec:cl_lds}. 
In Sec.~\ref{sub:bepn}, the binding energy per particle in nuclear matter is formulated with the TOFS framework.
{
In Sec.~\ref{sec:application}, the TOFS theory is applied for the study of the property of symmetric nuclear matter using the Argonne V4' \textit{NN} potential (central-force-type) with short-range repulsion~\cite{wiringa95}.
We compare our results with those of other theories (BHF etc.) as a sort of benchmarking purposes and confirm the validity of the TOFS theory.
}
Finally, the summary is presented in Sec.~\ref{sec:summary}.
In this paper, we treat only a symmetric nuclear matter system, but the present formulation {is} applicable to any infinite fermionic system such as neutron matter system and electron-gas system etc.

\section{Correlated wave function of nuclear matter in TOFS}
\label{sec:2}

In the TOFS formalism we take the following two types of the correlated nuclear matter wave function:~The first is the $N$th-order power series type and the second is the exponential type. 
The former is defined as  
\begin{eqnarray} 
&&\Psi_{N} = \left[ \sum_{k=0}^{N} \frac{1}{k!} F^{k} \right] \Phi_{0}, 
\label{eq:correlated_wf}\\
&&F = F_{S} + F_{D}.
\label{eq:correlation_fun}
\end{eqnarray} 
The exponential type is defined as the limiting case for Eq.~(\ref{eq:correlated_wf}) with $N \rightarrow \infty$,
\begin{eqnarray}
\Psi_{\rm ex} = \lim_{N \rightarrow \infty} \Psi_{N} = \lim_{N \rightarrow \infty} \left[ \sum_{k=0}^{N} \frac{1}{k!} F^{k} \right] \Phi_0 = \exp(F) \Phi_0.
\label{eq:correlated_wf_ex}
\end{eqnarray}
The uncorrelated wave function of the symmetric nuclear matter $\Phi_{0}$ is described by the Fermi  gas model, in which $A$ nucleons occupy up to the Fermi sea with the Fermi wave number $k_F$, 
\begin{eqnarray}
\Phi_{0} = \frac{1}{\sqrt{A!}}\ {\det}\ | \phi_{\gamma_1}(1) \phi_{\gamma_2}(2) \cdots \phi_{\gamma_A}(A)|.
\label{eq:femi_gas_wf}
\end{eqnarray}
The single-nucleon wave function $\phi_\gamma$ confined in a box with length $L$ and volume $\Omega=L^3$ is written as
\begin{eqnarray}
&&\phi_{\gamma_n}(n) = \phi_{\vc{k}_n}(\vc{r}_n)\ \chi_{m_{s_n}}(n)\  \xi_{m_{t_n}}(n),\nonumber\\
&&\phi_{\vc{k}_n}(\vc{r}_n) = \frac{1}{\sqrt{\Omega}} \exp ( i \vc{k}_n \cdot \vc{r}_n), 
\label{eq:spwf}
\end{eqnarray}
where $\gamma=(\vc{k},m_s,m_t)$ represents the quantum number of the single-nucleon wave function, and $\chi$ and $\xi$ are the spin and isospin functions, respectively.
Under the periodic boundary condition, the wave number vector $\vc{k}$ is discretized as $\vc{k} = \frac{2\pi}{L} (n_x, n_y, n_z)$, where $n_x$, $n_y$, and $n_z$ are integers. 
Then, the single-nucleon wave function $\phi_\gamma$ in Eq.~(\ref{eq:spwf}) is ortho-normalized.
The nucleon density of the nuclear matter is written
\begin{eqnarray}
\rho = \sum_{\gamma=1}^{A} |\phi_{\gamma_n}(\vc{r})|^2 = \frac{A}{\Omega}.
\end{eqnarray}
In infinite nuclear matter $(L \rightarrow \infty)$, the summation over $\gamma$ may be replaced as
\begin{eqnarray}
\sum_{\gamma=1}^{A}\ \rightarrow\ \frac{1}{4} \sum_{m_s=\pm 1/2} \sum_{m_t=\pm 1/2} \frac{4 \Omega}{(2\pi)^3} \int_{|k|\leq k_F} d\vc{k},
\label{eq:gamma_to_integral}
\end{eqnarray}
and then, we obtain $\rho = 2k_{F}^{3}/(3\pi^2)$. 

The correlation functions $F_S$ and $F_D$ in Eq.~(\ref{eq:correlation_fun}), which describe  the spin-isospin dependent central correlation and tensor correlation in nuclear matter, respectively, are defined as
\begin{eqnarray}
F_S &=& \frac{1}{2} \sum_{i  \not= j} f_S(i,j) = \frac{1}{2} \sum_{s=0}^{1}\sum_{t=0}^{1}\sum_{i \not= j} f_S^{(st)}(r_{ij})P^{(st)}_{ij}, 
\label{eq:fs}\\
F_D &=& \frac{1}{2} \sum_{i \not=j} f_D(i,j) = \frac{1}{2} \sum_{s=0}^{1}\sum_{t=0}^{1} \sum_{i \not= j}f_D^{(st)} (r_{ij}) r^2_{ij} S_{12}(i,j)P^{(st)}_{ij} \delta_{s1},
\label{eq:fd} 
\end{eqnarray}
where $S_{12}$ is the tensor operator, 
\begin{eqnarray}
&&S_{12}(i,j)= 3(\vc{\sigma}_i\cdot\hat{\vc{r}}_{ij})(\vc{\sigma}_j\cdot\hat{\vc{r}}_{ij}) - (\vc{\sigma}_i\cdot\vc{\sigma}_j)
\end{eqnarray}
with $\hat{\vc{r}}_{ij}=\vc{r}_{ij}/r_{ij}$ and $\vc{r}_{ij}=\vc{r}_{i} - \vc{r}_{j}$.
The operator $P^{(st)}_{ij}$ denotes the projection operator of the spin $s$ and isospin $t$ states of the $ij$-nucleon pair:~$P^{(st)}_{ij}=P^{(s)}_{ij} P^{(t)}_{ij}$ with 
\begin{eqnarray}
&&P^{(s)}_{ij}=\frac{1}{4}\left[ (2s+1)+(-1)^{s+1}(\vc{\sigma}_i \cdot \vc{\sigma}_j)\right], \\
&&P^{(t)}_{ij}=\frac{1}{4}\left[ (2t+1)+(-1)^{t+1}(\vc{\tau}_i \cdot \vc{\tau}_j)\right],
\end{eqnarray}
where the spin operators $\vc{\sigma}_i$ and $\vc{\sigma}_j$ (isospin operators $\vc{\tau}_i$ and $\vc{\tau}_j$) are for the particles {\it i} and {\it j}, respectively.
One may add the other type correlation functions such as the spin-orbit-type correlation function $F_{SO}$ into $F$ in Eq.~(\ref{eq:correlation_fun}) as needed, $F=F_{S}+F_{D}+F_{SO}$.

\section{Cluster expansion and linked diagram summation}\label{sec:cl_lds}

\subsection{Cluster expansion}\label{sub:cluster_expansion}

Let us consider the wave function $\Psi(\alpha)$ defined by
\begin{eqnarray} 
\Psi_{N}(\alpha) = \left[ \sum_{k=0}^{N} \frac{1}{k!} (\alpha F)^{k} \right] \Phi_{0}, 
\label{eq:cluster_Psi}
\end{eqnarray} 
where $\alpha$ is a real number, and $F$ and $\Phi_{0}$ stand for the Hermitian correlation operator of the $A$-body system in Eq.~(\ref{eq:correlation_fun}) and the Fermi-gas wave function in Eq.~(\ref{eq:femi_gas_wf}), respectively.
For an $M$-body operator (Hermite and translationally invariant) of the $A$-particle system 
\begin{eqnarray}
\hat{O}=\sum_{i_1 \not= i_2 \not= \cdots \not= i_M }^{A} \hat{O}(i_1,i_2,\cdots,i_M),
\end{eqnarray}
its expectation value with respect to the wave function $\Psi_{N}(\alpha)$ is presented as
\begin{eqnarray}
&&O_{N}(\alpha) 
\equiv \frac{\left\langle \Psi_{N}(\alpha) \left| \hat{O} \right|  \Psi_{N}(\alpha) \right\rangle}{\left\langle \Psi_{N}(\alpha) |  \Psi_{N}(\alpha) \right\rangle} 
= \frac{\displaystyle \sum_{n=0}^{2N} a_n \alpha^n}{\displaystyle \sum_{n=0}^{2N} b_n \alpha^n}
= \sum_{n=0}^{\infty} A_n \alpha^n, 
\label{eq:cluster_O_An}\\
&& A_n = \frac{a_n}{b_0} - \frac{b_1}{b_0}A_{n-1} - \cdots - \frac{b_{2N}}{b_0}A_{n-2N}, \\
&& a_n = \sum_{k=0}^{n} \frac{1}{(n-k)! k!} \left\langle \Phi_{0} \left| F^{n-k} \hat{O} F^{k} \right| \Phi_{0} \right\rangle, \label{eq:cluster_a_n}\\
&& b_n = \frac{2^n}{n!} \left\langle \Phi_{0} | F^{n} | \Phi_{0} \right\rangle, 
\label{eq:cluster_b_n}
\end{eqnarray}
where $A_i =0$ for $ i < 0 $, $b_0=1$, and $a_n=b_n=0$ for $n>2N$.
Setting $\alpha=1$ in Eq.~(\ref{eq:cluster_O_An}), we get the cluster expansion for the expectation of the operator $\hat{O}$,
\begin{eqnarray}
&&{\langle \hat{O} \rangle}_{N}
\equiv \frac{\left\langle \Psi_{N} \left| \hat{O} \right|  \Psi_{N} \right\rangle}{\left\langle \Psi_{N} |  \Psi_{N} \right\rangle} 
= \sum_{n=0}^{\infty} A_n,
\label{eq:ev_O_An}
\end{eqnarray}
where $\Psi_{N}$ is given in Eq.~(\ref{eq:correlated_wf}).

Here, it is instructive to consider the case of the limitation of $N \rightarrow \infty$ in Eq.~(\ref{eq:cluster_Psi})
\begin{eqnarray} 
\Psi_{\rm ex}(\alpha) = \lim_{N \rightarrow \infty} \Psi_{N}(\alpha) = \left[ \sum_{k=0}^{\infty} \frac{1}{k!} (\alpha F)^{k} \right] \Phi_{0} = \exp(\alpha F) \Phi_{0}.
\label{eq:cluster_Psi_infty}
\end{eqnarray} 
Then,  the expectation value of the operator $\hat{O}$ with respect to $\Psi_{\rm ex}(\alpha)$ is written     
\begin{eqnarray}
&&O_{\rm ex}(\alpha) 
\equiv \frac{\left\langle \Psi_{\rm ex}(\alpha) \left| \hat{O} \right|  \Psi_{\rm ex}(\alpha) \right\rangle}{\left\langle \Psi_{\rm ex}(\alpha) |  \Psi_{\rm ex}(\alpha) \right\rangle} 
= \frac{\displaystyle \sum_{n=0}^{\infty} a_n \alpha^n}{\displaystyle \sum_{n=0}^{\infty} b_n \alpha^n}
= \sum_{n=0}^{\infty} B_n \alpha^n, 
\label{eq:cluster_O_Bn}
\\
&& B_n = \frac{a_n}{b_0} - \sum_{k=1}^{n} \frac{b_{k}}{b_0}B_{n-k},
\label{eq:rf_Bn}
\end{eqnarray} 
where $a_{n}$ and $b_{n}$ are defined in Eqs.~(\ref{eq:cluster_a_n}) and (\ref{eq:cluster_b_n}), respectively.
It is noted that $B_n=A_n$ for $0 \le n \le N$.
Setting $\alpha=1$ in Eq.~(\ref{eq:cluster_Psi_infty}), the cluster expansion for the expectation of the operator $\hat{O}$ with respect to $\Psi_{\rm ex} = \exp(F) \Phi$ in Eq.~(\ref{eq:correlated_wf_ex}) is presented as
\begin{eqnarray}
&&{\langle \hat{O} \rangle}_{\rm ex}
\equiv \frac{\left\langle \Psi_{\rm ex} \left| \hat{O} \right| \Psi_{\rm ex} \right\rangle}{\left\langle \Psi_{\rm ex} | \Psi_{\rm ex} \right\rangle}
=   \frac{\left\langle \Phi_{0} \left| \exp(F^{\dagger}) \hat{O} \exp(F) \right| \Phi_{0} \right\rangle}{\left\langle \Phi_{0} |  \exp(F^{\dagger})  \exp(F) | \Phi_{0} \right\rangle}
= \sum_{n=0}^{\infty} B_n.
\label{eq:ev_O_infty}
\end{eqnarray}
In order to evaluate $a_n$ and $b_n$ as well as $A_n$ and $B_n$, we need to examine in more detail the analytical structure of $\left\langle \Phi_{0} | F^{n-k} \hat{O} F^{k} | \Phi_{0} \right\rangle$ and $\left\langle \Phi_{0} | F^{n} | \Phi_{0} \right\rangle$ in Eqs.~(\ref{eq:cluster_a_n}) and (\ref{eq:cluster_b_n}).
They have complex structures originating from the two facts:\ (1) Multi-body operators arise from the product operator $F^{n}$ together with $F^{n-k}\hat{O}F^{k}$, and (2) the Fermi-gas wave function $\Phi_{0}$ is totally antisymmetrized. 
We will discuss them in the next section.

\subsection{Linked diagram summation}\label{sub:lds}

The product operators, $F^{n}$ and $F^{n-k}\hat{O}F^{k}$, in Eqs.~(\ref{eq:cluster_a_n}) and (\ref{eq:cluster_b_n}) can be expanded as  the sum of the multi-body operators (see Appendix~\ref{app:A}).
In the case of $F^{n}$, this has the multi-body operators from the two-body to $2n$-body ones.
They are classified into the linked operator and unlinked operator (or separable operator).
For example, the product of the two symmetrized operators, $F_1$ and $F_2$, is expanded as follows:\ 
\begin{eqnarray}
&&F_1 F_2
= \left( \frac{1}{2} \sum_{i \not= j}^{A} f_{1}(ij) \right) \left( \frac{1}{2} \sum_{i \not= j}^{A} f_{2}(ij) \right), 
\label{eq:F1_F2_multi_operators_0}
\\
&& \hspace*{10mm}
= \frac{1}{2} \sum_{i \not= j}^{A} f_{1}(ij) f_{2}(ij) 
 + \sum_{i \not= j \not=k}^{A} f_{1}(ij) f_{2}(ik) 
 + \frac{1}{4} \sum_{i \not= j \not=k \not= l}^{A} f_{1}(ij) f_{2}(kl), 
\label{eq:F1_F2_multi_operators}
\\
&& \hspace*{10mm}
\equiv {\frac{1}{2} (12)^2} + 1(12)(13) + \frac{1}{4}(12)(34), 
\label{eq:F1_F2_multi_operators_1}
\\
&& \hspace*{10mm}
\equiv \wick{1}{ <1 {F_1} >1 F_2 } + \wick{1}{ <1 F_1 >1 {\hbox to-1.8ex{\vbox to0.7em{}}}{\kern1.2ex}}
   \wick{1}{ <1 F_2 >1 {\hbox to-1.8ex{\vbox to0.7em{}}}{\kern1.2ex}}\ .
\label{eq:F1_F2_multi_operators_3}
\end{eqnarray}
Here, the first and second terms in Eqs.~(\ref{eq:F1_F2_multi_operators}) and (\ref{eq:F1_F2_multi_operators_1}) are the two-body and three-body operators, respectively, which are called {\it linked} operators, because the operators $f_{1}(ij)f_{2}(ij)$ and $f_{1}(ij)f_{2}(ik)$ are linked (non-separable), respectively.
On the other hand, the third term, which is the four-body operator, is {\it unlinked}, because the operator $f_{1}(ij) f_{2}(kl)$ is separable.
The sum of the linked operators in $F_1 F_2$ is abbreviated as $\wick{1}{ <1 {F_1} >1 F_2 } \equiv \frac{1}{2}(12)^2 + 1(12)(13) \equiv \frac{1}{2}\sum_{i\not= j} f_1(ij) f_2(ij)  +\sum_{i\not=j\not=k} f_1(ij) f_2(ik)$, while that of the unlinked operators is given as $\wick{1}{ <1 F_1 >1 {\hbox to-1.8ex{\vbox to0.7em{}}}{\kern1.2ex}}
   \wick{1}{ <1 F_2 >1 {\hbox to-1.8ex{\vbox to0.7em{}}}{\kern1.2ex}}\ \equiv \frac{1}{4}(12)(34) \equiv \frac{1}{4} \sum_{i\not=j\not=k\not=l} f_1(ij) f_2(kl)$.
The multi-body operator expansion for the product of arbitrary symmetrized operators is formulated in Appendix~\ref{app:A}.

Let $Q_{M}$ be an arbitrary $M$-body operator appearing in the multi-body expansion of $F^{n}$ and $F^{n-k} \hat{O} F^{k}$, and be expressed in a product of non-separable operators of size $\alpha$, $Q_1$, $Q_2$, $\cdots$, and $Q_{\alpha}$,
\begin{eqnarray}
&&Q_{M}=\sum_{i_1 \not= i_2 \not= \cdots \not= i_M}^{A} Q_{M}(i_1,i_2,\cdots,i_M),\\ 
&&Q_{M}(1,2,\cdots,M) = Q_1(1,\cdots,n_{1}) Q_2(n_{1}+1,\cdots,n_{2}) \cdots Q_{\alpha}(n_{\alpha-1}+1,\cdots,n_{\alpha}),
\label{eq:Q1_Q_alpha}
\end{eqnarray}
with $1 \le \alpha \le M$, $n_0=0$, and ${n_\alpha}=M$.
Then, the matrix element of the operator $Q_{M}$ with respect to the Fermi-gas wave function $\Phi_{0}$ in Eq.~(\ref{eq:femi_gas_wf}) is written as
\begin{eqnarray}
{\langle Q_M \rangle} 
&=& {\left\langle \Phi_{0} \left| \sum_{i_1 \not= i_2 \not= \cdots \not= i_M}^{A} Q_{M}(i_1,i_2,\cdots,i_M) \right| \Phi_{0} \right\rangle}, \\
&=& \sum_{\gamma_1 \not= \gamma_2 \not= \cdots \not=\gamma_M}^{A} \left\langle \phi_{\gamma_1}(1) \phi_{\gamma_2}(2) \cdots \phi_{\gamma_M}(M) | Q_{M}(1,2,\cdots,M) | {\det} | \phi_{\gamma_1}(1) \phi_{\gamma_2}(2) \cdots \phi_{\gamma_M}(M) | \right\rangle.
\nonumber\\
\label{eq:QN}
\end{eqnarray}
In infinite nuclear matter, applying Eq.~(\ref{eq:gamma_to_integral}) to Eq.~(\ref{eq:QN}) instead of the summation over $\gamma$, the matrix element ${\langle Q_M \rangle}$ is presented as   
\begin{eqnarray}
{\langle Q_M \rangle} = A \times \rho^{M-1} \times \sum_{\beta} {\rm sgn}(\beta) \times \int d\vc{r}_{12} \cdots d\vc{r}_{1M}\ G_{\beta}(\{ \vc{r} \}), \hspace*{20mm}
\label{eq:QN_1}
\\
G_{\beta}(\{ \vc{r} \}) = \left[ \prod_{a=1}^{M} \left( \frac{4}{(2\pi)^{3} \rho} \int_{|\vc{k}_a| < k_F} d\vc{k}_{a} \right) \right]\  
\exp\left[ -i \sum_{a=1}^{M} \vc{k}_a \cdot \vc{r}_a \right] 
\label{eq:G_beta}
\hspace*{20mm}\nonumber \\
\hspace*{50mm} \times F_{\beta}(\{ \vc{r} \}) \times 
\exp\left[ i \sum_{a=1}^{M} \vc{k}_{\beta_{a}} \cdot \vc{r}_a \right],\hspace*{20mm}\\
F_{\beta}(\{ \vc{r}\}) = \frac{1}{4^{M}} 
\sum_{\substack{m_{s_1},\cdots,m_{s_M}\\m_{t_1},\cdots,m_{t_M}}} 
\left[ \prod_{a=1}^{M} \chi_{m_{s_{a}}(a)}\xi_{m_{t_{a}}(a)} \right]^{\dagger}
Q_M({1,\cdots,M})
\left[ \prod_{a=1}^{M} \chi_{m_{s_{\beta_{a}}}(a)}\xi_{m_{t_{\beta_{a}}}(a)} \right],\\
\beta=
\begin{pmatrix}
1 \cdots n_1 & n_1+1 \cdots n_2  & \cdots & n_{\alpha-1}+1 \cdots M \\
\beta_1 \cdots \beta_{n_1} & \beta_{n_1+1} \cdots \beta_{n_2} & \cdots & \beta_{n_{\alpha-1}+1} \cdots \beta_{M} \\
\end{pmatrix} 
\hspace*{30mm}
\nonumber \\
=
\begin{pmatrix}
 K_1 & K_2  & \cdots & K_{\alpha} \\
\beta(K_1) & \beta(K_2) & \cdots & \beta(K_{\alpha}) \\
\end{pmatrix},
\hspace*{57mm}
\label{eq:permutation_beta}
\end{eqnarray}
where we use the relation, $\prod_{a=1}^{M} d\vc{r}_a = d\vc{r}_{\rm cm} \prod_{a=2}^{M} d\vc{r}_{1a}$, and $\vc{r}_{\rm cm}$ denotes the center-of-mass coordinate of the $M$-body system, and $\vc{r}_{ab}$ is the relative coordinate between the $a$th and $b$th particles ($\vc{r}_{ab}=\vc{r}_{a} - \vc{r}_{b}$).  
It is remarked that the result of the integral $\frac{1}{\Omega} \int  d\vc{r}_{\rm cm} = 1$ is used in Eq.~(\ref{eq:QN_1}), because the integrand $G_{\beta}$ is in general expressed as the functional of the relative coordinates (see later).

\begin{figure}[t]
\begin{center}
\includegraphics*[width=0.8\hsize]{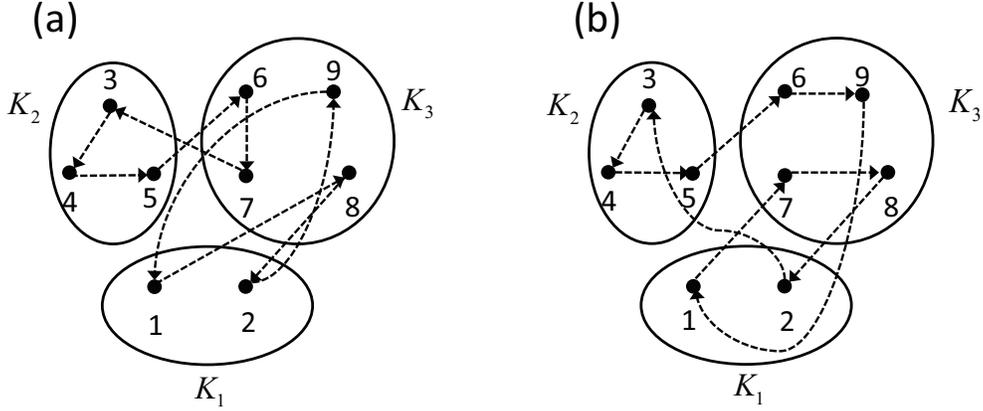}\\
\caption{
Examples of the diagrammatic representations for the linked permutation terms, (a)\ $\beta_2$ and (b)\ $\beta_3$, shown in Eq.~(\ref{eq:linked_ex}), where $M=9$, $\alpha=3$ with $K_{1}=(12)$, $K_{2}=(345)$, and $K_{3}=(6789)$. See the text.
}
\label{fig:permutation_term_examples}
\end{center}
\end{figure}

Here $\beta$ denotes the permutation for the numbers from $1$ to $M$ appearing in the Slater determinant, $ {\det}\ | \phi_{\gamma_1}(1) \cdots \phi_{\gamma_M}(M) | = \sum_{\beta} {\rm sgn}(\beta)\ \phi_{\gamma_{\beta_1}}(1) \cdots \phi_{\gamma_{\beta_M}}(M) $, in Eq.~(\ref{eq:QN}), and the numbers are grouped into the subgroups of size $\alpha$, $K_{1}=(1,\cdots,n_1)$, $K_{2}=(n_{1}+1,\cdots,n_2)$, $\cdots$, and $K_{\alpha}=(n_{\alpha-1}+1,\cdots,M)$ in accordance with the operator $Q_M$ expressed as the product of the non-separable operators of size $\alpha$ (see Eq.~(\ref{eq:Q1_Q_alpha})). 
The number of the permutation $\beta$ is $M!$.
They can be classified into the {\it linked } permutation terms and {\it unlinked} permutation terms according to the type of the permutation (\ref{eq:permutation_beta}).
The {\it unlinked} permutation term is defined as one that the permutation $\beta$ in Eq.~(\ref{eq:permutation_beta}) is presented in terms of the direct product of several sub-permutations with respect to $K_1$, $K_2$, $\cdots$, and $K_{\alpha} $.
For example, in the case of $M=9$ and $\alpha=3$ with $K_{1}=(12)$, $K_{2}=(345)$, and $K_{3}=(6789)$, the following permutation $\beta_{1}$ is unlinked,
\begin{eqnarray}
&&\beta_1 =
\begin{pmatrix}
12 & 345 & 6789 \\
21 & 456 & 7893 \\
\end{pmatrix}
=
\begin{pmatrix}
12 \\
21 \\
\end{pmatrix}
\begin{pmatrix}
345 & 6789 \\
456 & 7893 \\
\end{pmatrix}.
\end{eqnarray}
On the other hand, the {\it linked} permutation term is defined as one that $\beta$ is {\it not} presented in terms of the direct product of sub-permutations with respect to the sub-groups. 
For instance, let us consider the following two examples:\ 
\begin{eqnarray}
&&\beta_2 =
\begin{pmatrix}
12 & 345 & 6789 \\
89 & 456 & 7321 \\
\end{pmatrix},
\hspace*{5mm}
\beta_3 =
\begin{pmatrix}
12 & 345 & 6789 \\
73 & 456 & 9821 \\
\end{pmatrix}.
\label{eq:linked_ex}
\end{eqnarray}
In the case of  $\beta_2$, it has the permutation between $K_1$ and $K_3$, and that between $K_2$ and $K_3$, but no permutation between $K_1$ and $K_2$.
This type of permutation is {\it chain-type}.
The diagrammatic representation for $\beta_2$ is shown in Fig.~\ref{fig:permutation_term_examples}(a), where each permutation pair is connected by a broken line with an arrow.
On the other hand, in the case of $\beta_3$, there are the permutation between $K_1$ and $K_2$, that between $K_2$ and $K_3$, and that between $K_3$ and $K_1$.
This type of permutation is {\it ring-type}.
The diagrammatic representation for $\beta_3$ is shown in Fig.~\ref{fig:permutation_term_examples}(b).
It is noted that we can give the diagrammatic representation for any permutation term. 

If the operator $Q_M$ has no derivative terms, $G_{\beta}$ in Eq.~(\ref{eq:G_beta}) is given as
\begin{eqnarray}
&&G_{\beta}(\{ \vc{r} \}) = F_{\beta}(\{ \vc{r} \}) \times g_{\beta}(\{ \vc{r} \}),
\label{eq:QN_2}
\\
&&g_{\beta}(\{ \vc{r} \}) = 
\left[ \prod_{a=1}^{M} \left( \frac{4}{(2\pi)^3 \rho} \int_{{|\vc{k}_a|} \le k_F} d\vc{k}_a \right) \right]  
\exp\left[ -i \sum_{a=1}^{M} \vc{k}_a \cdot \vc{r}_a \right]
\exp\left[  i \sum_{a=1}^{M} \vc{k}_{\beta_{a}} \cdot \vc{r}_a \right],
\nonumber
\\
&&\hspace*{14mm}
=\prod_{a=1}^{M} h(k_F R_{a}(\beta)),
\label{eq:g_beta}
\\
&&
\vc{R}_a(\beta) = \sum_{j=1}^{M} \delta_{a,\beta_j} \vc{r}_j - \vc{r}_a,
\end{eqnarray}
where $h(k_F R)$ is defined as
\begin{eqnarray}
&&h(k_F R) = \frac{4}{(2\pi)^3 \rho} \int_{{\small |\vc{k}|} \le k_F} \exp(i\vc{k}\cdot\vc{R}) d\vc{k} = \frac{3j_1(k_F R)}{k_F R}.
\end{eqnarray}
Here, it is worthwhile to introduce a diagrammatic representation of $G_{\beta}$ given in Eq.~(\ref{eq:QN_2}).
We may represent it diagrammatically at the same manner as those in Fig.~\ref{fig:permutation_term_examples}:\ 
For each non-separable operator $Q_{i}(n_{i-1}+1,\cdots,n_{i})$ appearing in $F_{\beta}$ in Eq.~(\ref{eq:QN_2}), we can draw a circle enclosing the particle numbers belonging to $Q_i$ (denoted by black dots), and then each permutation pair in $\beta$ of $G_{\beta}$ is connected by a broken line with an arrow.
Since the operator $Q_i$ is non-separable, the particle numbers in the circle are regarded to be connected to each other. 
If all of the circles are connected by the broken lines, the diagram is called {\it linked}. 
This diagram is represented in a similar way to those shown in Figs.~\ref{fig:permutation_term_examples}(a) and (b).
On the other hand, if one or more circles are not connected by broken lines, this diagram is called {\it unlinked}.

In the case that the permutation term $\beta$ is linked, the diagram of $G_{\beta}$ is linked, even though the operator $Q_M$ is separable.
Then the integral $\int d\vc{r}_{12} \cdots d\vc{r}_{1M}\ G_{\beta}(\{ \vc{r} \})$ in Eq.~(\ref{eq:QN_1}) for the linked diagram is not divergent. 
However, in the unlinked diagram, the integral is proportional to $\Omega$ or higher power of $\Omega$, where $\Omega$ denotes the volume of infinite system.
Then its integral is divergent in infinite system. 
The reasons are given as follows:\ For the linked permutation term $\beta$, all of $\{ \vc{R}_{a}(\beta) \}_{a=1,\cdots,M}$ in Eq.~(\ref{eq:g_beta}) are nonzero, because the permutation term $\beta$ is linked.  
They are described by the linear combinations of the independent relative coordinates of size $M-1$ (for example, $\vc{r}_{12}$, $\vc{r}_{13}$, $\cdots$, $\vc{r}_{1M}$).
Then $g_{\beta}(\{ \vc{r} \})$ is expressed as a function of the independent relative coordinates of size $M-1$.
In addition, $F_{\beta}(\{ \vc{r} \})$ is also given as a function of the independent relative coordinates of size $M-1$, since the operator $Q_M$ is translationally invariant.
In this case each broken line in the linked diagram such as those in Figs.~\ref{fig:permutation_term_examples}(a) and (b) corresponds to any one of $\{ h(k_F R_{a}(\beta)) \}_{a=1,2,\cdots,M}$, and the particle numbers in each circle are connected because of the non-separable operators $Q_i$ with $i=1,\cdots,\alpha$.
As a result, the integral $\int d\vc{r}_{12} \cdots d\vc{r}_{1M}\ G_{\beta}(\{ \vc{r} \})$ is not divergent in infinite system.
We refer to it as {\it linked integral}. 

On the other hand, in the case of the unlinked permutation term  $\beta$, the number of the independent relative coordinates in $g_{\beta}(\{ \vc{r} \})$ is reduced by one or more from the size of $M-1$, and therefore some circles disconnected to other circles appear in the diagrammatic representation of $G_{\beta}(\{ \vc{r} \})$, although the particle numbers in $Q_i$ are connected inside their own circle.
Then, the integral $\int d\vc{r}_{12} \cdots d\vc{r}_{1M}\ G_{\beta}(\{ \vc{r} \})$ is proportional to $\Omega^{m}$, where $m$ ($\ge 1$) is the number of the independent relative coordinates reduced from the size of $M-1$.
Thus its integral is divergent in infinite system. 
We call it {\it unlinked integral}.

The above discussions hold in the case that $Q_{M}$ has derivative terms such as the kinetic energy operator and spin-orbit force etc.
It is noted that the kinetic energy operator of the {\it i}th particle, $-\frac{\hbar^2}{2m}\frac{\partial^2}{\partial \vc{r}_{i}^{2}}$, can be treated as the translationally invariant operator in the infinite nuclear matter.  
In this case, $G_\beta(\{ \vc{r} \})$ is in general expressed as a sum of several terms, $G_\beta(\{ \vc{r} \}) = \sum_{p} G_\beta^{(p)}(\{ \vc{r} \})$, which are originated from the character of the derivative terms in the operators, and may have some derivative terms of $h(k_F R_a)$.
The topological aspect of the diagrammatic representation for each $G_\beta^{(p)}(\{ \vc{r} \})$, however, is determined by the character of the permutation term $\beta$.
Thus, the diagrammatic representation for $G_\beta(\{ \vc{r} \}) = \sum_{p} G_\beta^{(p)}(\{ \vc{r} \})$ is linked, if $\beta$ is the linked permutation term, and vice versa.
 
As a consequence of the properties of the linked and unlinked diagrams together with the multi-body expansion of the product operator,  the matrix element of the operator $F^{n_{1}}\hat{O}F^{n_{2}}$ $(n_1, n_2 = 0, 1, 2, \cdots)$ with respect to $\Phi_{0}$, including the case of $\hat{O}=1$, can be divided into the two terms, 
\begin{eqnarray}
{\langle \Phi_{0} | F^{n_1} \hat{O} F^{n_2} | \Phi_{0} \rangle} = {\langle \Phi_{0} | F^{n_1} \hat{O} F^{n_2} | \Phi_{0} \rangle}_{\rm c} + {\langle \Phi_{0} | F^{n_1} \hat{O} F^{n_2} | \Phi_{0} \rangle}_{\rm dis}.   
\label{eq:c_dis_FOF}
\end{eqnarray}
The first term on the right side, called {\it linked matrix element} (or connected one), denotes the sum of all the linked integrals in the matrix element ${\langle \Phi_{0} | F^{n_1} \hat{O} F^{n_2} | \Phi_{0} \rangle}$, and the second term, called {\it unlinked matrix element} (or disconnected one), is that of all the unlinked integrals in the matrix element.
As proved in Appendix~\ref{app:C} with help from the results given in Appendices~\ref{app:A} and \ref{app:B}, the matrix elements of the operators $F^{n}$ ($n \ge 1$) and $F^{n_{1}}\hat{O}F^{n_{2}}$ $(n_1, n_2 = 0, 1, 2, \cdots$, and $\hat{O}\not= \hat{1})$  with respect to $\Phi_0$ are written as the following recurrence formulae,  
\begin{eqnarray}
&&{\langle \Phi_{0} | F^{n} | \Phi_{0} \rangle} = \sum_{k=0}^{n-1}
\frac{(n-1)!}{k!\ (n-k-1)!}\ {\langle \Phi_{0} | F^{n-k} | \Phi_{0}  \rangle}_{\rm c}\ {\langle \Phi_{0} |  F^{k} | \Phi_{0}  \rangle}, \\
&&{\langle \Phi_{0} | F^{n_1} \hat{O} F^{n_2} | \Phi_{0} \rangle} \nonumber \\ 
&&\hspace*{1mm}
= \sum_{k=0}^{n_1+n_2} \sum_{\substack{{k_1,k_2}\\k_1+k_2=k}} 
\frac{n_{1}!}{k_{1}!\ (n_{1} -k_{1})!}\ \frac{n_{2}!}{k_{2}!\ (n_{2} -k_{2})!}
{\langle \Phi_{0} | F^{k_1} \hat{O} F^{k_2} | \Phi_{0} \rangle_{\rm c}} {\langle \Phi_{0} | F^{n_1+n_2-k} | \Phi_{0} \rangle}, \nonumber \\
\label{eq:rc_mt_FOF}
\end{eqnarray}
where ${\langle \Phi_{0} | F^{0} | \Phi_{0} \rangle} = 1$.  

Here, we evaluate the summation of the linked diagrams in $a_n$ in Eq.~(\ref{eq:cluster_a_n}) and that of the unlinked ones, $(a_n)_{\rm c}$ and $(a_n)_{\rm dis}$, respectively, with use of Eqs.~(\ref{eq:c_dis_FOF}) $\sim$ (\ref{eq:rc_mt_FOF}),
\begin{eqnarray}
&&a_n = (a_n)_{\rm c} + (a_n)_{\rm dis},\\
&&(a_n)_{\rm c} = \sum_{\substack{n_1,n_2\\{n_1+n_2=n}}} \frac{1}{{n_1}!\  {n_2}!}\ {\left\langle \Phi_{0} \left| F^{n_1}\hat{O}F^{n_2} \right| \Phi_{0} \right\rangle}_{\rm c},
\label{eq:linked_a_n}
\\
&&(a_n)_{\rm dis} = \sum_{\substack{n_1,n_2\\{n_1+n_2=n}}} \frac{1}{{n_1}!\  {n_2}!}\ {\left\langle \Phi_{0} \left| F^{n_1}\hat{O}F^{n_2} \right| \Phi_{0} \right\rangle}_{\rm dis}, \nonumber\\
&&\hspace*{12mm}= \sum_{p=1}^{n} \frac{2^p}{p!}\ {\left\langle \Phi_{0} | F^p | \Phi_{0} \right\rangle}\  \sum_{q=0}^{n-p} \frac{1}{(n-p-q)!\ q!}\ {\left\langle \Phi_{0} \left| F^{n-p-q}\hat{O}F^{q} \right| \Phi_{0} \right\rangle}_{\rm c}, \nonumber \\
&&\hspace*{12mm}= \sum_{p=1}^{n} b_p \times (a_{n-p})_{\rm c},
\label{eq:(a_n)_dis}
\end{eqnarray}
where $b_n$ is defined in Eq.~(\ref{eq:cluster_b_n}).
From the recurrence formula for $B_n$ in Eq.~(\ref{eq:rf_Bn}), we can obtain the following important result,
\begin{eqnarray}
B_n = \frac{a_n}{b_0} - \sum_{k=1}^{n} \frac{b_{k}}{b_0}B_{n-k} = (a_n)_{\rm c}.
\label{eq:Bn_fin} 
\end{eqnarray}
This means that all the terms of the unlinked diagrams in $a_n$ cancel out exactly.
Then, from Eq.~(\ref{eq:ev_O_infty}), the expectation value of $\hat{O}$ with respect to the exponential correlated wave function $\Psi_{ex}$ defined in Eq.~(\ref{eq:correlated_wf_ex}) is given as the sum of the linked diagrams,
\begin{eqnarray}
&&{\langle \hat{O} \rangle}_{\rm ex}
= \frac{\left\langle \Psi_{\rm ex} \left| \hat{O} \right| \Psi_{\rm ex} \right\rangle}{\left\langle \Psi_{\rm ex} | \Psi_{\rm ex} \right\rangle}
= \sum_{n=0}^{\infty}
\sum_{\substack{n_1,n_2\\{n_1+n_2=n}}} \frac{1}{{n_1}!\  {n_2}!}\ {\left\langle \Phi_{0} \left| F^{n_1}\hat{O}F^{n_2} \right| \Phi_{0} \right\rangle}_{\rm c}.
\label{eq:ev_O_infty_result}
\end{eqnarray}

Finally we will discuss the case that the $N$th-order power series correlated wave function $\Psi_{N}$ in Eq.~(\ref{eq:correlated_wf}) is used to evaluate the expectation value of $\hat{O}$.
Taking into account the fact that $\Phi_N$ is the $N$th-order polynomial with respect to the correlation function $F$, one may take the following expression as an approximation of ${\langle \hat{O} \rangle}_{\rm ex}$ in Eq.~(\ref{eq:ev_O_infty_result}),
\begin{eqnarray}
&&{\langle \hat{O} \rangle}_{N}
= \frac{\left\langle \Psi_{N} \left| \hat{O} \right|  \Psi_{N} \right\rangle}{\left\langle \Psi_{N} |  \Psi_{N} \right\rangle} 
\simeq
\sum_{n_1=0}^{N} \sum_{n_2=0}^{N} \frac{1}{{n_1}!\  {n_2}!}\ {\left\langle \Phi_{0} \left| F^{n_1}\hat{O}F^{n_2} \right| \Phi_{0} \right\rangle}_{\rm c}.
\label{eq:ev_O_finite}
\end{eqnarray}
We call it the $N$th-order approximation for the expectation value of the operator $\hat{O}$ in the TOFS framework.
In the limit of $N \rightarrow \infty$, it converges to ${\langle \hat{O} \rangle}_{\rm ex}$ definitely.

\section{Binding energy per nucleon in nuclear matter}
\label{sub:bepn}

In this section we will discuss the method of calculating the binding energy per nucleon in nuclear matter within the present framework on the basis of the results in Sec.~\ref{sec:cl_lds}.
The Hamiltonian of the nuclear matter is expressed as a sum of the kinetic energies, two-body interactions, and three-body interactions,
\begin{eqnarray}
\mathcal{H} = \sum_{i=1}^{A} t_i + \frac{1}{2} \sum_{i  \not= j }^{A} v_{ij} + \frac{1}{6} \sum_{i \not= j \not= k}^{A} V_{ijk}.  
\label{eq:hamiltonian}
\end{eqnarray}

First we consider the binding energy per particle in nuclear matter, $B_{\rm ex}$, with use of the exponential correlated wave function $\Psi_{\rm ex}$ in Eq.~(\ref{eq:correlated_wf_ex}).
From Eqs.~(\ref{eq:ev_O_infty}) and (\ref{eq:Bn_fin}), it is presented as
\begin{eqnarray}
&&-B_{\rm ex} = \frac{1}{A} \frac{\left\langle \Phi |  \exp(F^{\dagger}) \mathcal{H} \exp(F)  | \Phi \right\rangle}{\left\langle \Phi | \exp(F^{\dagger}) \exp(F)  | \Phi \right\rangle}
= \frac{1}{A} \sum_{n=0}^{\infty} (E_n)_{\rm c},
\label{eq:B_ex}
\\
&&(E_n)_{\rm c} = \sum_{\substack{n_1,n_2\\{n_1+n_2=n}}} \frac{1}{{n_1}!\  {n_2}!}\ {\left\langle \Phi_0 | F^{n_1}\mathcal{H}F^{n_2} | \Phi_0 \right\rangle}_{\rm c},
\end{eqnarray}
where ${\left\langle \Phi_0 | F^{n_1}\mathcal{H}F^{n_2} | \Phi_0 \right\rangle}_{\rm c}$ is the summation of the linked integrals in the matrix element of ${\left\langle \Phi_0 | F^{n_1}\mathcal{H}F^{n_2} | \Phi_0 \right\rangle}$, defined in Eq.~(\ref{eq:Bn_fin}), and $(E_n)_{\rm c}$ stands for the $n$th-order cluster energy.
The explicit expressions of $(E_n)_{\rm c}$ with $n=0,1,2,3$ are presented as
\begin{eqnarray}
\begin{split}
(E_0)_{\rm c} &= {\left\langle \Phi_0 | \mathcal{H} | \Phi_0 \right\rangle}_{\rm c} = {\left\langle \Phi_0 | \mathcal{H} | \Phi_0 \right\rangle}, \nonumber\\
(E_1)_{\rm c} &= {\left\langle \Phi_0 | F\mathcal{H} + \mathcal{H}F | \Phi_0 \right\rangle}_{\rm c}, \nonumber\\
(E_2)_{\rm c} &= {\left\langle \Phi_0 \left| \frac{1}{2!} F^2\mathcal{H} + F\mathcal{H}F + \frac{1}{2!} \mathcal{H} F^2 \right| \Phi_0 \right\rangle}_{\rm c}, \nonumber\\
(E_3)_{\rm c} &= {\left\langle \Phi_0 \left| \frac{1}{3!} F^3\mathcal{H} + \frac{1}{2!}F^2\mathcal{H}F + \frac{1}{2!} F\mathcal{H} F^2 + \frac{1}{3!} \mathcal{H}F^3 \right| \Phi_0 \right\rangle}_{\rm c}.
\end{split}
\end{eqnarray}
It is noted that each integral of the linked diagram is proportional to the number $A$ (see Eq.~(\ref{eq:QN_1})), and therefore $B_{\rm ex}$ becomes $A$-independent or $\rho$-dependent.
When the product operator $F^{n_1}\mathcal{H}F^{n_2}$ is expanded into the multi-body operators, it is expressed as the sum of the multi-body operators from the two-body to $(2n_1+2n_2+3)$-body operators with $n_1+n_2 \ge 1$. 
For example, $(E_2)_{\rm c}$ is presented as the sum of the matrix elements from the 2-body to 7-body operators.  
It is remarked that all the linked diagrams can be classified into the two classes, `\textit{simple}' and  `\textit{composite}', and the former can be decomposed into two groups, `\textit{nodals}' and `\textit{elementary}', following the case of the FHNC framework~\cite{fantoni72,fantoni74,fantoni78,pandharipande79}.  

In taking the $N$th-order power series wave function $\Psi_{N}$ in Eq.~(\ref{eq:correlated_wf}) as the correlated wave function of nuclear matter, the following expression should be taken as an approximation for $B_{\rm ex}$, in accordance with the results in Sec.~\ref{sec:cl_lds},
\begin{eqnarray}
&&{-B}_{N}
= \frac{1}{A}\,\frac{\left\langle \Psi_{N} \left| \mathcal{H} \right|  \Psi_{N} \right\rangle}{\left\langle \Psi_{N} |  \Psi_{N} \right\rangle} 
\simeq
\frac{1}{A}\,\sum_{n_1=0}^{N} \sum_{n_2=0}^{N} \frac{1}{{n_1}!\  {n_2}!}\ {\left\langle \Phi_{0} \left| F^{n_1}\mathcal{H}F^{n_2} \right| \Phi_{0} \right\rangle}_{\rm c},
\label{eq:BE_linked}
\end{eqnarray}
where only the linked diagrams in the matrix element are evaluated.
We notice that the right side in Eq.~(\ref{eq:BE_linked}) is independent of $A$ or depends on only $\rho$, and converges definitely to $-B_{\rm ex}$ in Eq.~(\ref{eq:B_ex}) in the limit of $N \rightarrow \infty$.
Since the Hamiltonian $\mathcal{H}$ in Eq.~(\ref{eq:hamiltonian}) has one-body, two-body, and three-body operators, the product operator $F^{n_1}\mathcal{H}F^{n_2}$ is expressed as the summation from the one-body to $(2n_1+2n_2+3)$-body operators. 
We refer to evaluating $B_{N}$ in Eq.~(\ref{eq:BE_linked}) as the $N$th-order TOFS calculation. 

The explicit expressions of $-B_N$ with $N=1,2$ are given as
\begin{eqnarray}
&&-B_{N=1} = \frac{1}{A}\,\left[{\langle \Phi_0 | \mathcal{H} | \Phi_0 \rangle}_{\rm c} +  {\langle \Phi_0 | F\mathcal{H} + \mathcal{H}F | \Phi_0 \rangle}_{\rm c} +  {\langle \Phi_0 | F\mathcal{H}F | \Phi_0 \rangle}_{\rm c}\,\right], 
\label{eq:1st_TOFS_cal}
\\
&&-B_{N=2} = \frac{1}{A} \left[ {\langle \Phi_0 | \mathcal{H} | \Phi_0 \rangle}_{\rm c} +  {\langle \Phi_0 | F\mathcal{H} + \mathcal{H}F | \Phi_0 \rangle}_{\rm c} +  {\left\langle \Phi_0 \left| \frac{1}{2!}F^2\mathcal{H} + F\mathcal{H}F + \frac{1}{2!} \mathcal{H}F^2 \right| \Phi_0 \right\rangle}_{\rm c} \right. \nonumber \\
&&\hspace*{25mm}\left.+\ {\left\langle \Phi_0 \left| \frac{1}{2!}F^2\mathcal{H}F + \frac{1}{2!} F\mathcal{H}F^2 \right| \Phi_0 \right\rangle}_{\rm c}
+ {\left\langle \Phi_0 \left| \frac{1}{{2!}^2}F^2\mathcal{H}F^2 \right| \Phi_0 \right\rangle}_{\rm c}\,\right],
\label{eq:1st_2nd_TOFS_cal}
\end{eqnarray}
where $F=F_S + F_D$ are given in Eq.~(\ref{eq:correlation_fun}).

The present TOFS framework is in contrast to the framework of TOAMD~\cite{myo15,myo17_1,myo17_2,myo17_3,myo17_4,myo17_5} and HM-AMD~\cite{myo17_6,lyu17}, which  treat the finite nuclei with the bare interaction among nucleons.
In the TOAMD and HM-AMD formalism, the correlated nuclear wave function is described by the arbitrary power series function with respect to $F_S$ and $F_D$ multiplied to the uncorrelated AMD wave function. 
In nuclear matter, however, the arbitrary power series function with respect to $F_S$ and $F_D$ is not allowed for the correlated nuclear matter wave function, because in this case the unlinked diagrams are not canceled out completely.
One needs to employ the present TOFS framework for describing the correlated power series nuclear matter wave function, which can take into account only the linked diagrams.    

In the TOFS framework, the radial parts of the correlation functions in Eqs.~(\ref{eq:fs}) and ($\ref{eq:fd}$) are expanded in terms of the Gaussian functions,
\begin{eqnarray}
&&f_S^{(st)}(r) = \sum_{\mu}C^{(st)}_{S,\mu} \exp\left[-a^{(st)}_{S,\mu} r^2\right],\\
&&f_D^{(st)}(r) = \sum_{\mu}C^{(st)}_{D,\mu} \exp\left[-a^{(st)}_{D,\mu} r^2\right].
\end{eqnarray}
Here, $C^{(st)}_{S,\mu}$ and $a^{(st)}_{S,\mu}$ together with $C^{(st)}_{D,\mu}$ and $a^{(st)}_{D,\mu}$ are the variational parameters.
They are determined so as to minimize the energy per nucleon in nuclear matter.
It is noted that the Gaussian correlation functions bring about simplification and numerical stabilization for evaluating the matrix elements of many-body operators in nuclear matter calculation~\cite{av4_yamada}.  
In the actual numerical calculations~\cite{av4_yamada}, the values of the size parameters, $a^{(st)}_{S,\mu}$  and $a^{(st)}_{D,\mu}$, are appropriately chosen by considering the range of the nuclear force and the Fermi wave number $k_F$, and only the expansion coefficients are taken as the variational parameters, i.e.~they are determined from the following conditions:
\begin{eqnarray}
\frac{\partial B_{N}}{\partial C^{(st)}_{S,\mu}}=0,\ \ \ \ \ 
\frac{\partial B_{N}}{\partial C^{(st)}_{D,\mu}}=0. 
\label{eq:system_eq}
\end{eqnarray}
The solutions to the system of equations $(\ref{eq:system_eq})$ for $C^{(st)}_{S,\mu}$ and $C^{(st)}_{D,\mu}$ give the variational minimization of the energy per particle in nuclear matter $(-B_{N})$ in Eq.~(\ref{eq:BE_linked}).

In the 1st-order TOFS calculation, $-B_{N=1}$ in Eq.~(\ref{eq:1st_TOFS_cal}) is described by the quadratic form in terms of the expansion coefficients, $C^{(st)}_{S,\mu}$ and $C^{(st)}_{D,\mu}$.
These expansion coefficients are obtained by solving {\it a system of linear equations} with respect to $C^{(st)}_{S,\mu}$ and $C^{(st)}_{D,\mu}$, which are derived from Eq.~(\ref{eq:system_eq}) with $B_{N}=B_{N=1}$. 
This is a remarkable aspect in the 1st-order TOFS calculation.
In the 2nd-order TOFS calculation (see Eq.~(\ref{eq:1st_2nd_TOFS_cal})), the variational conditions in Eq.~(\ref{eq:system_eq}) give a system of non-linear equations in terms of $C^{(st)}_{S,\mu}$ and $C^{(st)}_{D,\mu}$. 
These equations are solved easily and numerically . 
 
In the $N$th-order TOFS calculation, the effects of many-body correlations originating from the correlated Hamiltonian $\left({\sum_{n=0}^N \frac{1}{n!} {F}^n}\right) \mathcal{H} \left({\sum_{n=0}^N \frac{1}{n!} F^n}\right)$ in Eq.~(\ref{eq:BE_linked}) can be taken into account as the follows:~We evaluate all the contribution from one-body to $(2N+3)$-body terms arising from the product operator $\left({\sum_{n=0}^N \frac{1}{n!} {F}^n}\right) \mathcal{H} \left({\sum_{n=0}^N \frac{1}{n!} F^n}\right)$.
Then we can include all the `\textit{elementary}' and `\textit{nodals}' diagrams together with the `\textit{composite}' ones which appear in calculating the matrix element in Eq.~(\ref{eq:BE_linked}).   
This point is different from the variational method based on the hypernetted chain summation techniques (VCS or FHNC-SOC) by Pandharipande et al.~\cite{pandharipande79,akmal98}.
In the VCS, the many-body correlations are taken into account through the `\textit{nodals}' diagrams presented by the single operator chain (SOC) approximation together with their `\textit{composite}' diagrams, while the `\textit{elementary}' diagrams are neglected.
For the finite nuclei the computations using the correlated basis function theory and Fermi hypernetted chain theory~\cite{co94,arias07} has been performed for estimating the effect of the `\textit{elementary}' diagrams.
They say that the effect of the `\textit{elementary}' diagrams is not negligible in the binding energies of the ground states of $^{16}$O and $^{40}$Ca.
Thus, the quantitative estimation for the effect of the `\textit{elementary}' diagrams seems to be of  importance in nuclear matter, presumably of importance in dense nuclear matter. 

It is instructive to discuss the difference between the present TOFS theory and the CC theory~\cite{baardsen13,hagen14,kummel78,barlett81}.
In the CC formalism, the correlated wave function is taken as the exponential type,
\begin{eqnarray}
\Psi_{\rm CC} = \exp(\hat{S}) \Phi_{0},
\end{eqnarray}
where $\Phi_{0}$ is the uncorrelated free Fermi vacuum, and the cluster operator $\hat{S}$ is defined as the sum 
\begin{eqnarray}
\hat{S} = \sum_{m=1}^{A} \hat{S}_m
\end{eqnarray}
of $m$-particle $m$-hole excitation operators.
This correlated wave function is similar to that in the present TOFS framework in the case of the limitation of $N \rightarrow \infty$, shown in Eq.~(\ref{eq:correlated_wf_ex}). 
However, in the CC theory, the energy per particles in nuclear matter is evaluated as 
\begin{eqnarray}
-B_{\rm CC} = \frac{1}{A} {\langle \Phi_{0} | \exp(-\hat{S}) \mathcal{H}\exp(\hat{S}) | \Phi_{0} \rangle},
\label{eq:B_CC}
\end{eqnarray}
where the cluster operator $\hat{S}$ is obtained from the corresponding set of CC amplitude equations.
It is noted that in the TOFS framework the correlated Hamiltonian, $\exp(F^\dagger)\mathcal{H}\exp(F)$, in Eq.~(\ref{eq:B_ex}) is {\it Hermitian}, together with the {\it Hermitian} norm operator, $\exp(F^{\dagger}) \exp(F)$, while in the CC framework that is {\it non-Hermitian}, $\exp(-S) \mathcal{H} \exp(S)$, shown in Eq.~(\ref{eq:B_CC}). 
This point arises major difference between the TOFS and CC formalisms.

\section{Application of the TOFS theory to nuclear matter}
\label{sec:application}

The 1st-order TOFS theory is applied for the study of the property of symmetric nuclear matter using the Argonne V4' (AV4') \textit{NN} potential (central-force-type) with short-range repulsion~\cite{wiringa95}.
This calculation is performed as a sort of benchmarking purposes.
The AV4' Hamiltonian with the $A$-body system is written as 
\begin{eqnarray}
\mathcal{H} 
= \sum_{i=1}^{A}\,\left( -\frac{\hbar^2}{2m} \vc{\nabla}^2_i \right) + \frac{1}{2} \sum_{i \not=j}^{A} \left[ \sum_{s=0}^{1} \sum_{t=0}^{1} v_{C}^{(st)}(r_{ij}) P^{(st)}_{ij} \right],
\label{eq:hamiltonian_central}
\end{eqnarray}
where $m$ denotes the nucleon mass, and $P^{(st)}_{ij}=P^{(s)}_{ij} P^{(t)}_{ij}$ stands for the spin-isospin projection operator.
The AV4' potential reproduces the binding energy of the deuteron in the $^{3}S_{1}$ channel, and the \textit{NN} phase shifts of the  $^{1}S_{1}$, $^{3}S_{1}$, and $^{1}P_{1}$ channels are reasonably reproduced up to energies of about 350 MeV, while those of $^{3}S_{3,2}$ and $^{3}P_{0,1,2}$ are not well because of no tensor coupling etc.
The repulsive strength of the potential at $r=0$ is as strong as a couple of GeV for each spin-isospin channel.


\begin{table}[t]
\caption{Density dependence of the energy per particle for symmetric nuclear matter calculated by the 1st-order TOFS method with the AV4' potential, compared with those of the BHF approach~\cite{baldo12}. 
The energy is given in the unit of MeV.}
\begin{center}
\begin{tabular}{crrrr}
\hline
$\rho$ [fm${^{-3}}$]  & \ 0.05 & \ 0.10 & \ 0.17 & \ 0.20 \\
\hline
~~1st order TOFS~~  & \ $-8.5 $\  & \ $-16.1$\  & \ $-26.8$\   & \ $-32.9$\  \\
BHF~\cite{baldo12} &  $-11.4$ & $-17.7$ & $-26.4$ & $-29.7$ \\
\hline
\end{tabular}
\end{center}
\label{tab:1}
\end{table}

The density dependence of the energy per particle for symmetric nuclear matter, $E/A=-B_{N=1}$ in Eq.~(\ref{eq:1st_TOFS_cal}), with the 1st-order TOFS calculation is shown in Table~\ref{tab:1}.
We found that the results of the present calculation are reasonably reproduced, compared with those of the BHF approach~\cite{baldo12}, although a couple of MeV attraction is deficient in lower density ($\rho=0.05$ fm$^{-3}$) and a few MeV overbinding is seen at higher density ($\rho=0.20$ fm$^{-3}$).
As noted in Sec.~\ref{sec:Introduction}, the density dependence of $E/A$ for symmetric nuclear matter on the calculated methods such as BBG, SCGF, AFDMC, and FHNC is almost identical to that of BHF in the case of the AV4' potential, but the slight difference among them (a few MeV) emerges in the region of $\rho \ge \rho_0=0.17$~fm$^{-3}$ (see Fig.~3 in Ref.~\cite{baldo12}). 
Therefore, the overbinding by a few MeV at $\rho=0.20$ fm$^3$ is likely to be within the difference among the calculated methods.
However, the shortage of the attraction at the lower density may come from the fact that the present calculation is the lowest order one in the TOFS framework. 
The 2nd-order TOFS calculation in Eq.~(\ref{eq:1st_2nd_TOFS_cal}) is considered to be able to recover the small shortage.
The detailed discussions are provided in Ref.~\cite{av4_yamada}.
At any rate, the present calculated results confirm the reliability of the TOFS theory.


\section{Summary}
\label{sec:summary}

We have developed the new formalism for treating the nuclear matter with the bare interaction among nucleons, called tensor optimized Fermi sphere (TOFS) method.
In this formalism based on Hermitian form, the $N$th-order power series correlated wave function is used for describing the nuclear matter, $\Psi_{N} = [\sum_{k=0}^{N} (1/n!) F^k] \Phi_{0}$, together with the exponential type, $\Psi_{\rm ex} = \exp(F) \Phi_{0}$, where the correlation function $F$ is presented as the sum of the spin-isospin dependent central correlation ($F_S$) and tensor correlation ($F_D$) in nuclear matter, and $\Phi_{0}$ denotes the uncorrelated Fermi-gas wave function.
$\Psi_{\rm ex}$ corresponds to the limiting case of $\Psi_{N}$ with $N \rightarrow \infty$.
It was proved that the energy per particle in nuclear matter, $-B_{\rm ex}$, for $\Psi_{\rm ex}$ can be expressed as the sum of the linked diagrams with use of the cluster expansion with respect to $F$, and the unlinked diagrams are canceled out at each order of the cluster expansion.
Based on these results, the formula of the energy per particle in nuclear matter, $-B_{N}$, for $\Psi_{N}$ was given as an approximation of $B_{\rm ex}$.  
This formula of $B_{N}$ is useful to perform actual numerical calculations of nuclear matter. Evaluating $B_{N}$ is called the $N$th-order TOFS calculation, where $B_N$ converges to $B_{\rm ex}$ in limiting case of $N \rightarrow \infty$.

In the $N$th-order TOFS calculation, the correlation functions, $F_S$ and $F_D$, are expanded into the Gaussian functions, and the expansion coefficients $C_{\mu}$'s are treated as the variational parameters for evaluating $B_{N}$, although the ranges of the Gaussian functions are properly chosen by taking into account the range of the nuclear force and Fermi wave number $k_F$ (or nuclear matter density $\rho$).  
In evaluating $B_N$, we make consideration all the matrix elements of the many body operators coming from the product of operators, $F^{n_1}\mathcal{H}F^{n_2}$, with $ 0 \le n_1, n_2 \le N$, where $F^{n_1}\mathcal{H}F^{n_2}$ with the Hamiltonian of nuclear matter $\mathcal{H}$ appears in calculating $B_N$.
When $\mathcal{H}$ includes up to the three-nucleon force, the 1-body to $(2n_1+2n_2+3)$-body operators come out from the multi-operator expansion of $F^{n_1}\mathcal{H}F^{n_2}$, because of the two-body operator for $F_S$ and $F_D$.
Then, the value of $B_N$ is variationally determined with respect to $C_{\mu}$'s.
For example, in the 1st-order TOFS calculation ($N=1$), $B_{N=1}$ is determined by solving a system of \textit{linear} equations with respect to $C_{\mu}$'s, which are derived from the variational conditions for $B_{N=1}$ in terms of $C_{\mu}$'s, because $B_{N=1}$ is presented as a quadratic form of $C_{\mu}$'s.
On the other hand, in the case of $N \ge 2$, the expansion coefficients $C_{\mu}$'s in the case of $B_N$ are determined by solving a system of \textit{nonlinear} equations for $C_{\mu}$'s.  
We have discussed the differences between the TOFS and VCS (or FHNC-SOC) theories together with the CC theory.

{
We have applied the 1st-order TOFS theory for the study of the property of symmetric nuclear matter using the AV4' \textit{NN} potential (central-force-type) with short-range repulsion.
This calculation was made as a sort of benchmarking purposes.
It was found that the density dependence of the energy per nucleon, $E(\rho)/A$, is reasonably reproduced, in comparison with other methods such as the BHF approach etc.
This result confirms the reliability of the TOFS theory, although the 1st-order TOFS calculation is the first step to go to more sophisticated and higher-order calculations of symmetric nuclear matter. 
As noted in Sec.~\ref{sec:Introduction}, the significant dependence of $E(\rho)/A$ on the calculated methods (BHF, BBG, SCGF, FHNC, AFDMC) has been reported in the case of the AV6', AV8', and AV18 \textit{NN} potentials (having non central components such as tensor force etc.), even in the lower density including the normal density.
Therefore, it is interesting to study $E(\rho)/A$ using AV6', AV8', and AV18 potentials within the frame of the present TOFS theory.
}

In this paper, we have treated only the symmetric nuclear matter system.
The present formulation {is} applicable to any infinite fermionic system such as neutron matter system and electron-gas system etc.
{
In future it is highly hoped that the TOFS theory is widely used to perform calculations of the properties of the strongly correlated fermion systems.}

\section*{Acknowledgments}

We are grateful to Profs.~T.~Myo, K.~Ikeda, H.~Horiuchi, and H.~Toki for careful reading and useful discussions.
This work was partially supported by the JSPS KAKENHI Grants No.~26400283 and 18K03629.



\newpage

\appendix
\section{Multi-body operator expansion of operator product}\label{app:A}


First we consider the product of the two-body operators of size $n$ in $A$-body system, which is composed of $F_1$, $F_2$, $\cdots$, and $F_n$. 
It can be expressed as the sum of the multi-body operators from the two-body to $2n$-body operator, 
\begin{eqnarray}
F_1 F_2 \cdots F_n 
&=& \left( \frac{1}{2} \sum_{i \not= j}^{A} f_{1}(ij) \right) \left( \frac{1}{2} \sum_{i \not= j}^{A} f_{2}(ij) \right) \cdots \left( \frac{1}{2} \sum_{i \not= j}^{A} f_{n}(ij) \right), 
\label{eq:F1F2...Fn}
\\
&=& \frac{1}{2} \sum_{i_1 \not= i_2}^{A} f_1(i_1,i_2) f_2(i_1,i_2) \cdots f_n(i_1,i_2)  + \cdots \nonumber \\ 
&&\hspace*{10mm}+\  \frac{1}{2^n} \sum_{i_1 \not= i_2 \cdots \not= i_{2n}}^{A} f_1(i_1,i_2) f_2(i_3,i_4) \cdots f_n(i_{2n-1},i_{2n}), 
\label{eq:F1F2...Fn_1}\\
&\equiv& \frac{1}{2}(12)^n + 1(12)^{n-1}(34) + \cdots + \frac{1}{2^n}(12)(34)\cdots(2n-1,2n).
\label{eq:simple_F1F2...Fn_1}
\end{eqnarray}
In the last line, we use an irreducible expression for describing the configuration of the multi-body operator, where we take the following three rules:~(i)~In each parenthesis, left-side number is the smallest one as like (12) instead of (21), (ii)~in the product of parentheses, the parenthesis with smaller numbers is preferable more to the left side as like (12)(34) instead of (34)(12), and (iii)~the smaller number is given in the parentheses as much as possible as like (12)(13) instead of (12)(23), which are identical.
The general expression of each irreducible multi-body operator appearing in Eqs.~(\ref{eq:F1F2...Fn_1}) and (\ref{eq:simple_F1F2...Fn_1}) can be abbreviated as
\begin{eqnarray}
S \times {\sum_{\{i \}}^{A}}\, f_1(i_1,i_2) f_2(i_3,i_4) \cdots f_n(i_{2n-1},i_{2n}) 
\equiv S \times (i_1,i_2) (i_3,i_4) \cdots (i_{2n-1},i_{2n}), 
\label{eq:mbo}
\end{eqnarray}
where $S$ stands for the symmetry factor, and $\sum_{\{i \}}$ means a summation over with $\{ i \} = \{i_1,i_2,\cdots,i_{2n}\}$ under some restriction on $\{ i \}$.

\begin{figure}[t]
\begin{center}
\includegraphics*[width=0.5\hsize]{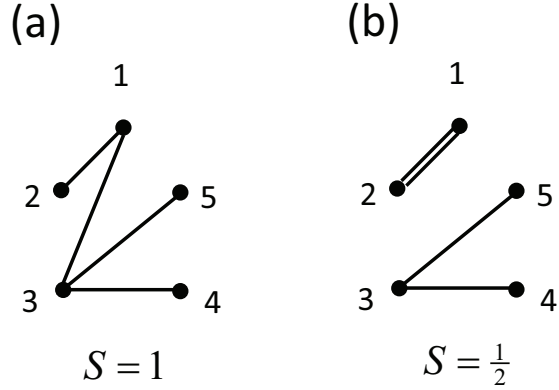}\\
\caption{
Diagrammatic representations for the five-body operators:\ (a) linked operator, $1(12)(34)(35)(13)$, and (b) unlinked one, $\frac{1}{2}(12)(34)(35)(12)$. They appear in the multi-body operator expansion of $F_1 F_2 F_3 F_4$, and $S$ stands for the symmetry factor. See the text.
}
\label{fig:linked_unlinked_operator_examples}
\end{center}
\end{figure}

Here, we define linked operator and unlinked operator as follows.
Let us consider Eq.~(\ref{eq:mbo}) as the $m$-body operator. 
Then we may connect particles $i_1$ and $i_2$ with a line, and in the same manner we connect particles $i_3$ and $i_4$, $\cdots$, and particles $i_{2n-1}$ and $i_{2n}$, respectively, with a line, and finally we get a diagram.
As a result, if all the $m$ particles are connected by lines in the diagram, we call the operator {\it linked}, while we call it {\it unlinked} if the lines cannot connect all the $m$ particles, in other words, the diagram is disconnected.
For example, the five-body operator  $1(12)(34)(35)(13)$, which is contained within the expansion of $F_1 F_2 F_3 F_4$, is linked, while $\frac{1}{2}(12)(34)(35)(12)$ is unlinked:\ 
\begin{eqnarray}
&&1(12)(34)(35)(13) = 1 \times \sum_{i_{1} \not= i_{2} \not= i_{3} \not= i_{4} \not= i_{5}} f_{1}(i_1,i_2) f_{2}(i_3,i_4) f_{3}(i_3,i_5) f_{4}(i_1,i_3), \nonumber \\
&&\frac{1}{2}(12)(34)(35)(12) = \frac{1}{2} \times \sum_{i_{1} \not= i_{2} \not= i_{3} \not= i_{4} \not= i_{5}} f_{1}(i_1,i_2) f_{2}(i_3,i_4) f_{3}(i_3,i_5) f_{4}(i_1,i_2). \nonumber
\end{eqnarray}
They are diagrammatically shown in Fig.~\ref{fig:linked_unlinked_operator_examples}(a) and (b), respectively.

The multi-body operators in Eq.~(\ref{eq:simple_F1F2...Fn_1}) can be classified into the linked and unlinked ones.
The results for the operator $F_1 F_2$ are given in Eqs.~(\ref{eq:F1_F2_multi_operators_0}) $\sim$ (\ref{eq:F1_F2_multi_operators_3}).
As for the operator $F_1 F_2 F_3$, we write down as follows,
\begin{eqnarray}
&&F_1 F_2 F_3 
= \wick{1}{ <1 F_1 >1 {\hbox to-1.8ex{\vbox to0.7em{}}}{\kern1.2ex}}{F_2}{F_3} 
+\wick{1}{ <1 {F_1} >1 {F_2} } {F_3} 
+\wick{1}{ <1 {F_1} {F_2} >1 {F_3}}
+ \wick{11}{ <1 {F_1} <2 {F_2} >1 {\hbox to-1.2ex{\vbox to1.em{}}}{\kern1.2ex} 
  >2 {F_3}}, 
\label{eq:F1F2F3} \\
&&\hspace*{14mm}= \wick{1}{ <1 F_1 >1 {\hbox to-1.8ex{\vbox to0.7em{}}}{\kern1.2ex}}\wick{1}{ <1 F_2 >1 {\hbox to-1.8ex{\vbox to0.7em{}}}{\kern1.2ex}} \wick{1}{ <1 F_3 >1 {\hbox to-1.8ex{\vbox to0.7em{}}}{\kern1.2ex}}
+\wick{1}{ <1 F_1 >1 {\hbox to-1.8ex{\vbox to0.7em{}}}{\kern1.2ex}}\wick{1}{ <1 {F_2} >1 F_3 } 
+\wick{1}{ <1 {F_1} >1 {F_2} } \wick{1}{ <1 F_3 >1 {\hbox to-1.8ex{\vbox to0.7em{}}}{\kern1.2ex}} 
+\wick{21}{ <1 {F_1} <2 {F_2} >2 {\hbox to-0.6ex{\vbox to0.7em{}}}{\kern1.2ex} >1 {F_3}}
+ \wick{11}{ <1 {F_1} <2 {F_2} >1 {\hbox to-1.2ex{\vbox to1.em{}}}{\kern1.2ex} 
  >2 {F_3}}, 
\label{eq:F1F2F3_1} \\
&&\wick{1}{ <1 F_1 >1 {\hbox to-1.8ex{\vbox to0.7em{}}}{\kern1.2ex}}{F_2}{F_3} 
= \frac{1}{2}(12) \times \left\{ \frac{1}{2} (34)^2 + 1(34)(35) + \frac{1}{4} (34)(56) \right\}, \nonumber \\
&&\wick{1}{ <1 {F_1} >1 {F_2} } {F_3}
= \left\{ \frac{1}{2} (12)^2 + 1 (12)(13) \right\} \times \left( \frac{1}{2}(45) \right), \nonumber \\
&&\wick{1}{ <1 {F_1} {F_2} >1 {F_3}}
= \frac{1}{2}(12) \left( \frac{1}{2}(34) \right) (12) + 1(12) \left( \frac{1}{2}(34) \right) (15), \nonumber \\
&&\wick{11}{ <1 {F_1} <2 {F_2} >1 {\hbox to-1.2ex{\vbox to1.em{}}}{\kern1.2ex} 
  >2 {F_3}}
= \frac{1}{2}(12)^3 + 1(12)^2(13) + 1(12)(13)(12) + 1(12)(13)^2 + 1(12)(13)(23) \nonumber \\
&&\hspace*{15mm} +\ 1(12)(13)(14) + 1(12)(13)(24) + 1(12)(13)(34) + 1 (12)(34)(13). \nonumber
\end{eqnarray}
Here, $\wick{1}{ <1 F_1 >1 {\hbox to-1.8ex{\vbox to0.7em{}}}{\kern1.2ex}}{F_2}{F_3}$ with $F_2F_3 = \wick{1}{ <1 {F_2} >1 F_3 } + \wick{1}{ <1 F_2 >1 {\hbox to-1.8ex{\vbox to0.7em{}}}{\kern1.2ex}}
   \wick{1}{ <1 F_3 >1 {\hbox to-1.8ex{\vbox to0.7em{}}}{\kern1.2ex}}\ $ represents the set of the multi-body operators in which particle numbers are unlinked between $F_1$ and $F_2 F_3$, and $\wick{1}{ <1 {F_1} >1 {F_2} } {F_3}$ $(\wick{1}{ <1 {F_1} {F_2} >1 {F_3}})$ does in which the operators $F_1$ and $F_2$ ($F_1$ and $F_3$) are linked but they are unlinked with $F_3$ ($F_2$). 
Other notations are given as follows:\ $\wick{21}{ <1 {F_1} <2 {F_2} >2 {\hbox to-0.6ex{\vbox to0.7em{}}}{\kern1.2ex} >1 {F_3}} = \wick{1}{ <1 {F_1} {F_2} >1 {F_3}}$, $\wick{1}{ <1 {F_1} {F_2} >1 {F_3}} = \wick{21}{ <1 {F_1} <2 {F_2} >2 {\hbox to-0.6ex{\vbox to0.7em{}}}{\kern1.2ex} >1 {F_3}} $, $\wick{1}{ <1 F_1 >1 {\hbox to-1.8ex{\vbox to0.7em{}}}{\kern1.2ex}}\wick{1}{ <1 F_2 >1 {\hbox to-1.8ex{\vbox to0.7em{}}}{\kern1.2ex}} \wick{1}{ <1 F_3 >1 {\hbox to-1.8ex{\vbox to0.7em{}}}{\kern1.2ex}} = \frac{1}{8}(12)(34)(56)$, and $\wick{1}{ <1 F_1 >1 {\hbox to-1.8ex{\vbox to0.7em{}}}{\kern1.2ex}}\wick{1}{ <1 {F_2} >1 F_3 }= \frac{1}{2}(12) \{ \frac{1}{2}(34)^2 + 1(34)(35) \} $ etc. 

The generalized decomposition of the product operator $F_1 F_2 \cdots F_{n}$ in Eq.~(\ref{eq:F1F2...Fn}) into the linked and unlinked operators can be performed by using the following recurrence formula
\begin{eqnarray}
&&F_1 F_2 \cdots F_n = \ Q^{(n)}_{1}(1) + \sum_{k=2}^{n}\ \ \sum_{2 \le i_2 < i_3 < \cdots < i_k \le n} Q^{(n)}_{k}(1,i_2,i_3,\cdots,i_k),
\label{eq:expansion_F1_Fn}
\end{eqnarray}
where $Q^{(n)}_{k}$\ $(k=1,\cdots,n)$  are defined as
\begin{eqnarray}
Q^{(n)}_{1}(1) &=& \wick{1}{ <1 F_1 >1 {\hbox to-1.8ex{\vbox to0.7em{}}}{\kern1.2ex}}{F_2}{\cdots}{F_n}, \nonumber \\
Q^{(n)}_{2}(1,i_{2}) &=& \wick{1}{ <1 {F_1} {\cdot\cdot\cdot} >1 {F_{i_{2}} }} {\cdots} {F_n}, \nonumber \\
Q^{(n)}_{3}(1,i_{2},i_{3}) &=& \wick{11}{ <1 {F_1} {\cdot\cdot\cdot}  <2 {F_{i_{2}}} >1 {\hbox to-1.2ex{\vbox to1.em{}}}{\kern1.2ex} {\cdot\cdot\cdot} >2 {F_{i_{3}}}} \cdots {F_{n}},  \nonumber \\
&&\hspace*{-20mm} \cdots\cdots\cdots\cdots\cdots\cdots \nonumber \\
Q^{(n)}_{k}(1,i_2,i_3,\cdots,i_k) &=& \wick{1111}{ <1 {F_{1}} {\cdot\cdot\cdot} <2 {F_{i_{2}}} {\cdot\cdot\cdot} >2 {F_{i_{3}}} {\cdot\cdot\cdot} {\hbox to-1.2ex{\vbox to1.em{}}}{\kern1.2ex} <3 {\cdot\cdot\cdot} >3 {\hbox to-1.2ex{\vbox to1.em{}}}{\kern1.2ex} {\cdot\cdot\cdot}  <4 {\cdot\cdot\cdot} >4 {\hbox to-1.2ex{\vbox to1.em{}}}{\kern1.2ex} {\cdot\cdot\cdot}>1 {F_{i_{k}} }} {\cdots} {F_n}, \nonumber\\
&&\hspace*{-20mm} \cdots\cdots\cdots\cdots\cdots\cdots \nonumber \\
Q^{(n)}_{n}(1,2,3,\cdots,n) &=& \wick{11111}{ <1 {F_{1}} <2 {F_{2}} <4 {F_{3}} {\cdot\cdot\cdot} >4 {\hbox to-1.2ex{\vbox to1.em{}}}{\kern1.2ex} >2  {\hbox to-1.2ex{\vbox to1.em{}}}{\kern1.2ex} <4 {\cdot\cdot\cdot} {\cdot\cdot\cdot\cdot\cdot\cdot} >4 {\hbox to-1.2ex{\vbox to1.em{}}}{\kern1.2ex} <3 {F_{n-2}} >3 {F_{n-1}} {\hbox to-1.2ex{\vbox to0.8em{}}}{\kern1.2ex} >1 {F_{n} }}. \nonumber
\end{eqnarray}
Here, $Q^{(n)}_{1}(1)$ denotes the multi-body operator group in which particle numbers are unlinked between $F_1$ and $F_2 F_3 \cdots F_n$. 
In a similar fashion,  $Q^{(n)}_{k}(1,i_2,i_3,\cdots,i_k)$ is that the $k$ operators, $F_1$, $F_{i_2}$, $\cdots$, $F_{i_k}$ are linked but they are unlinked with the remaining operators, and $Q^{(n)}_n(1,2,3,\cdots,n)$ is that where all operators $F_1$, $F_{2}$, $\cdots$, $F_{n}$, are linked.
The proof of Eq.~(\ref{eq:expansion_F1_Fn}) is given with use of the mathematical induction for instance.

From Eq.~(\ref{eq:expansion_F1_Fn}), one can get the expansion formula of the operator product $F_1F_2 \cdots F_n$ in terms of the linked operators,
\begin{eqnarray}
&&{F_1 F_2 \cdots F_n } \nonumber \\
&&\hspace*{5mm}=\ \sum_{ \{k \} } \ 
\left[ \prod_{1 \le p_1 \le n}  \left( \wick{1}{ <1 F_{p_1} >1 {\hbox to-1.8ex{\vbox to0.7em{}}} }\ \ \right)^{k^{(1)}_{p_1}} \right]
\left[ \prod_{1 \le p_1 < p_2 \le n} \left( \wick{1}{<1 {F_{p_1}} >1 {F_{p_2} }}\   \right)^{k^{(2)}_{p_1 p_2}} \right] \nonumber \\
&&\hspace*{10mm}\times\ 
\left[ \prod_{1 \le p_1 < p_2 < p_3 \le n} \left( \wick{11}{ <1 {F_{p_1}} <2 {F_{p_{2}}} >1 {\hbox to-1.2ex{\vbox to1.em{}}}{\kern1.2ex} >2 {F_{p_{3}}}} \right)^{k^{(3)}_{p_1 p_2 p_3}} \right] \nonumber \\
&&\hspace*{10mm}\times\ \cdots\ \times
\left[ \prod_{1 \le p_1 < p_2 \cdots < p_{n-1} \le n} \left( \wick{11111}{ <1 {F_{p_1}} <2 {F_{p_2}} <4 {F_{p_3}} {\cdot\cdot\cdot} >4 {\hbox to-1.2ex{\vbox to1.em{}}}{\kern1.2ex} >2  {\hbox to-1.2ex{\vbox to1.em{}}}{\kern1.2ex} <4 {\cdot\cdot\cdot} {\cdot\cdot\cdot\cdot\cdot\cdot} >4 {\hbox to-1.2ex{\vbox to1.em{}}}{\kern1.2ex} <3 {F_{p_{n-3}}} >3 {F_{p_{n-2}}} {\hbox to-1.2ex{\vbox to0.8em{}}}{\kern1.2ex} >1 {F_{p_{n-1}} }} \right)^{k^{(n-1)}_{p_1 p_2 \cdots p_{n-1}}} \right]  \nonumber \\
&&\hspace*{10mm}
\times\ 
\left( \wick{11111}{ <1 {F_{1}} <2 {F_{2}} <4 {F_{3}} {\cdot\cdot\cdot} >4 {\hbox to-1.2ex{\vbox to1.em{}}}{\kern1.2ex} >2  {\hbox to-1.2ex{\vbox to1.em{}}}{\kern1.2ex} <4 {\cdot\cdot\cdot} {\cdot\cdot\cdot\cdot\cdot\cdot} >4 {\hbox to-1.2ex{\vbox to1.em{}}}{\kern1.2ex} <3 {F_{n-2}} >3 {F_{n-1}} {\hbox to-1.2ex{\vbox to0.8em{}}}{\kern1.2ex} >1 {F_{n} }} \right)^{k^{(n)}_{1 2 3 \cdots, n-2, n-1, n}},
\label{eq:expansion_F1_Fn_connected_operator}
\end{eqnarray}
where the summation of $\sum_{ \{ k \} }$ over $\{ k \} =\{ k^{(1)}_{1}, k^{(1)}_{2}, \cdots, k^{(n)}_{1 2 \cdots n} \}$ is taken under the condition,
\begin{eqnarray}
&&k^{(1)}_{\beta} + {\sum_{p_1 < \beta} k^{(2)}_{p_1 \beta}} + {\sum_{\beta < p_1} k^{(2)}_{\beta p_1} } \nonumber \\
&&\hspace*{6mm}+\ {\sum_{p_1 <p_2 < \beta} k^{(3)}_{p_1 p_2\beta}} + {\sum_{p_1 < \beta <p_2 } k^{(3)}_{p_1 \beta p_2}} + {\sum_{ \beta < p_1 < p_2 } k^{(3)}_{\beta p_1  p_2}} 
\nonumber \\
&&\hspace*{6mm}+\cdots\cdots\cdots \nonumber \\
&&\hspace*{6mm}+\ {\sum_{p_1 < \cdots < p_{m-1} < \beta} k^{(m)}_{p_1 \cdots p_{m-1} \beta}}
+\ {\sum_{p_1 < \cdots < p_{m-2} < \beta < p_{m-1}} k^{(m)}_{p_1 \cdots p_{m-2} \beta p_{m-1}}} \nonumber \\
&&\hspace*{20mm}+\cdots\cdots\ + {\sum_{\beta < p_1 < \cdots < p_{m-1}} k^{(m)}_{\beta p_1 \cdots p_{m-1}}} \nonumber \\
&&\hspace*{6mm}+\cdots\cdots\cdots \nonumber \\
&&\hspace*{6mm}+\ k^{(n)}_{1 2 \cdots n} =1  \ \ \ {\rm with}\ {\beta=1,\cdots,n}.
\end{eqnarray}

In the above procedure, we restrict that all the operators $F_1$, $F_2$, $\cdots$, $F_{n}$ are two-body one.
However, the present procedure can be applicable in the case that the operators include one-body operator and three-body operator such as the kinetic energy operator $\sum_{i} t_i$ and three-nucleon potential operator $\sum_{i < j < k} V_{ijk}$, respectively.
The proof is the same as those given in this section. 

\section{Linked- and unlinked-term expansion of Slater determinant}\label{app:B}

The Slater determinant of the $N$-particle system with the orthonormal single-particle wave functions $\{ \phi_{\gamma} \}$ is expressed as
\begin{eqnarray}
D^{1, 2, \cdots, N}_{1, 2, \cdots, N}
= {\det}\, \left|\phi_{1}(1) \phi_{2}(2)\ \cdots\ \phi_{N}(N)\right|,
\label{eq:det_m}
\end{eqnarray}
where $\gamma$ in $\phi_{\gamma}$ is the index representing the single-particle state. 
The subscripts in $D$ denote the particle indices, and the superscripts in $D$ represent the indices of the single-particle states.
Let us split the $N$ single-particle states into the sub-groups of size $\alpha$, $K_1$, $K_2$, $\cdots$, and $K_{\alpha}$, where the states from the $1$st state to the $n_1$th one are allocated to the $K_1$ group, those from the $(n_1+1)$th state to the $n_2$th one belong to the $K_2$ group, $\cdots$, and  those from the $(n_{\alpha-1}+1)$th state to the $n_{\alpha}$th one are allocated to the $K_{\alpha}$ group ($n_{\alpha}=N$) :\ 
\begin{eqnarray}
&&K_1\ =\ (\ 1\ 2\ \cdots\ n_1\ ), \nonumber \\
&&K_2\ =\ (\ n_1+1\ n_1+2\ \cdots\ n_2\ ), \nonumber \\
&&\cdots\cdots\cdots \nonumber \\
&&K_{\alpha}\ =\ (\ n_{\alpha-1}+1\ n_{\alpha-1}+2\ \cdots\ n_{\alpha}\ ). \nonumber 
\end{eqnarray}
Hereafter, the following notations are used for the Slater determinant in Eq.~(\ref{eq:det_m}),
\begin{eqnarray}
D^{1, 2, \cdots, N}_{1, 2, \cdots, N}
&=& \sum_{\sigma}\ {\rm sgn}(\sigma)\ \phi_{\sigma_1 1} \phi_{\sigma_2 2}\ \cdots \ \phi_{\sigma_N N}, \nonumber\\
&\equiv&  {\det} \left| \phi_{K_1} \phi_{K_2} \ \cdots \ \phi_{K_{\alpha}} \right|, 
\nonumber \\
&\equiv&  \left( {K_1} {K_2} \ \cdots \ {K_{\alpha}} \right) \label{eq:det_m_1},
\label{eq:slater_K}
\end{eqnarray}
where $i$ and $j$ in $\phi_{ij}$ stand for the  single-particle-state index and particle index, respectively, and $\phi_{K_p}$ is equal to $\phi_{n_{p-1}+1\, n_{p-1}+1} \phi_{n_{p-1}+2\, n_{p-1}+2} \cdots \phi_{n_p\, n_p}$ with $p=1,2,\cdots,\alpha$ and $n_0=0$.

With use of the Laplace expansion theorem, we can expand the Slater determinant with respect to the rows (single-particle-state indices) from the $1$st row to the $n_1$th one,   
\begin{eqnarray}
&& D^{1, 2, \cdots, N}_{1, 2, \cdots, N} = 
\sum_{ 
\substack{ 
{1 \le \sigma_1 < \sigma_2 < \cdots < \sigma_{n_1} \le N}\\ 
( \sigma_{n_{1}+1} < \sigma_{n_{1}+2} < \cdots < \sigma_{n_{\alpha}}=\sigma_N ) 
}} (-1)^{\sum_{i=1}^{n_1} \sigma_i + \frac{1}{2}n_{1}(n_{1}+1)}\,D^{\sigma_1, \cdots, \sigma_{n_1}}_{1, \cdots, n_1}\,D^{\sigma_{n_1+1}, \cdots, \sigma_{N}}_{n_1+1, \cdots, N}, 
\label{eq:laplace} \\
&&D^{\sigma_1,\cdots,\sigma_{n_1}}_{1,\cdots,n_1} = {\det} \left| \phi_{\sigma_1 1}\ \sim\ \phi_{\sigma_{n_1} n_1} \right| \equiv (K_1), \nonumber \\
&&D^{\sigma_{n_{1}+1},\cdots,\sigma_{N}}_{n_1+1,\cdots,N} = {\det} \left| \phi_{\sigma_{n_{1}+1} n_{1}+1}\ \cdots \phi_{\sigma_{N} N} \right|
\equiv (K_2 K_3\ \cdots\ K_{\alpha}),\nonumber \\
&&\sigma =
\begin{pmatrix}
1 \cdots n_1 & n_1+1 \cdots n_2  & \cdots & n_{\alpha-1}+1 \cdots N \\
\sigma_1 \cdots \sigma_{n_1} & \sigma_{n_1+1} \cdots \sigma_{n_2} & \cdots & \sigma_{n_{\alpha-1}+1} \cdots \sigma_{N} \\
\end{pmatrix} 
=
\begin{pmatrix}
 K_1 & K_2  & \cdots & K_{\alpha} \\
\sigma(K_1) & \sigma(K_2) & \cdots & \sigma(K_{\alpha}) \\
\end{pmatrix},
\nonumber
\\
\label{eq:signature}
\end{eqnarray}
where $\sigma$ denotes the permutation relevant to the indices of the single particle states. 
The Slater determinant of the $N$-particle system has the permutation terms of size $N!$, which can be classified into the {\it linked} permutation terms and {\it unlinked} permutation terms in accordance with the type of the permutation $\sigma$ shown in Eq.~(\ref{eq:signature}), as discussed below.

Let us define the {\it unlinked permutation term} as one that the permutation $\sigma$ in Eq.~(\ref{eq:signature}) is presented in terms of the direct product of several sub-permutations with respect to the sub-groups of size $\alpha$, $K_1$, $K_2$, $\cdots$, and $K_{\alpha}$.
For example, in the case of that the set of $\{\sigma_1, \sigma_2, \cdots, \sigma_{n_1}\}$ is equal to that of $\{1, 2, \cdots, n_1\}$ and the set of $\{\sigma_{{n_1}+1} \sigma_{{n_1}+2},\cdots,\sigma_N \}$ is equal to that of $\{ n_1+1, n_1+2, \cdots, N \}$, its permutation $\sigma$ is presented as
\begin{eqnarray}
\sigma=
\begin{pmatrix}
1\ \cdots\ n_1 \\
\sigma_1\ \cdots\ \sigma_{n_1} \\
\end{pmatrix}
\begin{pmatrix}
n_1+1\ \cdots\ N \\
\sigma_{n_1+1}\ \cdots \sigma_N \\
\end{pmatrix}
=
\begin{pmatrix}
K_1 \\
\sigma(K_1) \\
\end{pmatrix}
\begin{pmatrix}
K_2\ \cdots\ K_{\alpha} \\
\sigma(K_2)\ \cdots\ \sigma(K_{\alpha}) \\
\end{pmatrix}.
\label{eq:exmple_unlinked}
\end{eqnarray}
Then, the summation of the unlinked permutation terms satisfying the condition of Eq.~(\ref{eq:exmple_unlinked}) is given as the product of sub-determinants
\begin{eqnarray}
&&{\sum}' \ {\rm sgn}(\sigma_1\ \cdots\ \sigma_{n_1})\ {\rm sgn}(\sigma_{n_1+1} \ \cdots\ \sigma_{N})\ \phi_{\sigma_1 1}\ \cdots\ \phi_{\sigma_{n_1} {n_1}} \ \phi_{\sigma_{n_{1}+1} {n_{1}+1}}\ \cdots\ \phi_{\sigma_{N} {N}} \nonumber \\
&&\hspace*{10mm}=\ D^{1,\cdots,n_1}_{1,\cdots,n_1} \times D^{n_1+1,\cdots,N}_{n_1+1,\cdots,N} 
\equiv\ (K_1) \times (K_2 \cdots K_{\alpha}). \nonumber
\end{eqnarray}
This result is just the direct term in Eq.~(\ref{eq:laplace}).
The unlinked permutation terms in Eq.~(\ref{eq:det_m}) is in general expressed as the sum of the product of two or more sub-determinants.

On the other hand, the {\it linked permutation term} is defined as one which is {\it not} unlinked permutation term, in other words, as one that the permutation $\sigma$ in Eq.~(\ref{eq:signature}) is {\it not} presented as the direct product of sub-permutations with respect to the sub-groups of size $\alpha$, $K_1$, $K_2$, $\cdots$, and $K_{\alpha}$. 
The examples of the linked permutation terms are given in Eq.~(\ref{eq:linked_ex}) together with their diagrammatic representations (see Fig.~\ref{fig:permutation_term_examples}).

Each linked permutation term appearing in Eq.~(\ref{eq:det_m}) as well as their sum can {\it not} be expressed as the product of two or more sub-determinants. 
Hereafter, the following notation is used for the summation of the linked permutation terms in Eq.~(\ref{eq:det_m}),
\begin{eqnarray}
&&{\sum_{\rm linked}}' \ {\rm sgn}(\sigma_1 \cdots \sigma_{N})\ \phi_{\sigma_1 1} \cdots \phi_{\sigma_{N} {N}}\
\equiv \ \left(\ {\det}\ | \phi_{K_1} \cdots \phi_{K_{\alpha}}|\ \right) _{\rm c}
\equiv \ \left( K_1 \cdots K_{\alpha} \right) _{\rm c},
\label{eq:sum_of_linked_terms}
\end{eqnarray}
where the subscript `c' denotes the linked.
The summation of the linked permutation terms is obtained by subtracting the summation of the unlinked ones from  the determinant ${\det}\ |\phi_{K_1}\ \cdots\ \phi_{K_{\alpha}}| = \left( K_1\ \cdots\ K_{\alpha} \right) $.
For example,
\begin{eqnarray}
&&(K_1 K_2)_{\rm c} = {{\det} | \phi_{K_1} \phi_{K_2} |} - \left({\det} | \phi_{K_1} | \right) \left({\det} | \phi_{K_2} | \right) = (K_1 K_2) - (K_1)(K_2), \label{eq:(K1K2)_c} \\
&&(K_1 K_2 K_3)_{\rm c} = (K_1 K_2K_3) - (K_1) (K_2 K_3) - (K_1 K_2)_{\rm c} (K_3) - (K_1 K_3)_{\rm c} (K_2), \label{eq:(K1K2K3)_c}\\
&&\hspace*{1mm}=\  (K_1 K_2K_3) - \left[ (K_1) (K_2) (K_3) + (K_1 K_2)_{\rm c} (K_3) + (K_1 K_3)_{\rm c} (K_2) + (K_2 K_3)_{\rm c} (K_1) \right]. 
\label{eq:(K1K2K3)_c_1} 
\end{eqnarray}

The general manner of expanding the Slater determinant of the $N$-particle system into the linked and unlinked permutation sets with respect to the sub-groups of size $\alpha$, $K_1$, $K_2$, $\cdots$, $K_{\alpha}$, is written as follows:\ 
\begin{eqnarray}
D^{1,\cdots,N}_{1,\cdots,N} &=& {\det}\ | \phi_{K_1}\ \phi_{K_2}\ \cdots\ \phi_{K_{\alpha}} | \equiv (K_1 K_2 \cdots K_{\alpha}) \nonumber \\
&=& \ (K_1)\ (\overline{K_1}) \nonumber \\
&+& \sum_{2 \le p_{2} \le \alpha} (K_1 K_{p_2})_{\rm c}\ (\overline{K_1 K_{p_2}}) \nonumber \\
&+& \sum_{2 \le p_{2} < p_{3} \le \alpha} (K_1 K_{p_2} K_{p_3})_{\rm c}\ (\overline{K_1 K_{p_2}  K_{p_3}}) \nonumber \\
&+& \sum_{2 \le p_{2} < p_{3} < p_{4} \le \alpha} (K_1 K_{p_2} K_{p_3} K_{p_4})_{\rm c}\ (\overline{K_1 K_{p_2} K_{p_3} K_{p_4}}) \nonumber \\
&+& \cdots\cdots\cdots \nonumber \\
&+& \sum_{2 \le p_{2} < \cdots < p_{n-1} \le \alpha} (K_1 K_{p_2} \cdots K_{p_{\alpha-1}})_{\rm c}\ (\overline{K_1 K_{p_2} \cdots K_{p_{\alpha-1}}}) \nonumber \\
&+& 
{\sideset{}{'}\sum_{
\substack{
1 \le \sigma_1 < \sigma_2 \cdots < \sigma_{n_1} \le N \\ 
( \sigma_{n_{1}+1} < \sigma_{n_{1}+2} < \cdots < \sigma_{n_{\alpha}}=\sigma_N )}}}
\left(-1\right)^{\sum_{i=1}^{n_1}\sigma_{i}+\frac{1}{2}n_1(n_1+1)} D^{\sigma(K_1)}_{K_1} D^{\sigma(K_2),\cdots,\sigma(K_{\alpha})}_{K_2,\cdots, K_{\alpha}},\nonumber\\
\label{eq:det_expansion}
\end{eqnarray}
where $(\overline{K_1})$ denotes the determinant in which $K_1$ is removed from $(K_1 K_2\ \cdots\ K_{\alpha}) = {\det} | \phi_{K_1} \phi_{K_2}\ \cdots\  \phi_{K_{\alpha}} |$, i.e.\ $(\overline{K_1}) = (K_2 K_3 \cdots K_{\alpha}) = {{\det} | \phi_{K_2} \phi_{K_3} \cdots \phi_{K_{\alpha}} |}$, and $(\overline{K_1 K_3}) = (K_2 K_4 \cdots K_{\alpha}) = {\det} |\phi_{K_2} \phi_{K_4} \cdots \phi_{K_{\alpha}}|$, and so on.
The last term in Eq.~(\ref{eq:det_expansion}) stands for the total sum of the linked permutation terms, i.e.~$(K_1 K_2 \cdots K_{\alpha})_c$, expressed in Eq.~(\ref{eq:sum_of_linked_terms}), and $\sum'$ means to take summation under the following conditions:\  
\begin{eqnarray}
&&\sigma(K_1) \not\in \{ \tau(K_1) \}, \nonumber \\
&&\sigma(K_1)\sigma(K_{p_2}) \not\in \{ \tau(K_1 K_{p_2})_{\rm c} \},\ 2 \le  p_2 \le \alpha, \nonumber \\
&&\sigma(K_1)\sigma(K_{p_2})\sigma(K_{p_3}) \not\in \{ \tau(K_1 K_{p_2} K_{p_3})_{\rm c} \},\ 2 \le  p_2 < p_3  \le \alpha, \nonumber \\
&&\sigma(K_1)\sigma(K_{p_2})\sigma(K_{p_3})\sigma(K_{p_4}) \not\in \{ \tau(K_1 K_{p_2} K_{p_3} K_{p_4})_{\rm c} \},\ 2 \le  p_2 < p_3  < p_4 \le \alpha, \nonumber \\
&& \cdots\cdots\cdots\cdots\cdots\cdots \nonumber \\
&&\sigma(K_1)\sigma(K_{p_2})\cdots\sigma(K_{p_{\alpha-1}}) \not\in \{ \tau(K_1 K_{p_2} \cdots K_{p_{\alpha-1}})_{\rm c}  \},\ 2 \le  p_2 < \cdots < p_{\alpha-1}  \le \alpha, \nonumber 
\end{eqnarray}
where $\sigma(K_1) = (\sigma_1,\sigma_2,\cdots,\sigma_{n_1})$ etc.~are defined in Eq.~(\ref{eq:signature}).
$\{ \tau(K_1) \}$ expresses the set of all the permutations for $K_1=(1,2,\cdots,n_1)$ with the elements of $n_{1}!$, i.e.\ $\{ \tau(K_1) \} =\{ (1,2,\cdots,n_{1}-1,n_{1}), \cdots, (n_{1}, n_{1}-1, \cdots,2,1) \}$, which appear in the determinant $(K_1) = {\det}\, |\, \phi_{K_1}\,|$.
Therefore, the condition of $\sigma(K_1) \not\in \{ \tau(K_1) \}$ means that all the permutations appearing in $(K_1) = \det\, |\, \phi_{K_1}\, |$ are removed in the sum of $\sum'$. 
On the other hand, $\sigma(K_1)\sigma(K_{p_2}) = (\sigma_1,\cdots,\sigma_{n_1},\sigma_{n_{(p_2-1)}+1},\cdots,\sigma_{n_{p_2}})$, and $\{ \tau(K_1 K_{p_2})_{\rm c} \}$ denotes the set of all the linked permutations belonging to $(K_1 K_{p_2})_{\rm c} = ( {\det} | \phi_{K_1} \phi_{K_{p_2}}| )_{\rm c}$.
Thus, the condition of $\sigma(K_1)\sigma(K_{p_2}) \not\in \{ \tau(K_1 K_{p_2})_{\rm c} \}$ indicates that all the permutations appearing in $(K_1 K_{p_2})_{\rm c}$  are removed in the sum of $\sum'$. 
Other conditions are self-explanatory. 
One can derive Eqs.~(\ref{eq:(K1K2)_c})$\sim$(\ref{eq:(K1K2K3)_c_1}) from Eq.~(\ref{eq:det_expansion}).
The proof of Eq.~(\ref{eq:det_expansion}) is easily given on the basis of the Laplace expansion theorem in Eq.~(\ref{eq:laplace}). 

\section{Matrix element of operator product with Slater determinant}\label{app:C}


First we consider the matrix element of the product of non-separable operators with Slater determinant. 

Let $Q$ be a symmetric $N$-body operator in the $A$-body system expressed in the product of the non-separable operators of size $\alpha$, $Q_1$, $Q_2$, $\cdots$, and  $Q_{\alpha}$,
\begin{eqnarray}
&&Q=\sum_{i_1 \not= i_2 \not= \cdots \not= i_N}^{A} Q(i_1,i_2,\cdots,i_N), 
\label{eq:Q1_Qn}\\
&&Q(1,2,\cdots,N) = Q_1(1,\cdots,n_{1}) Q_2(n_{1}+1,\cdots,n_{2}) \cdots Q_{\alpha}(n_{\alpha-1}+1,\cdots,n_{\alpha}),
\end{eqnarray}
with $1 \le \alpha \le N$, $n_{0}=0$, and $n_{\alpha}=N$.
Here, we consider the matrix element of $Q$ with respect to the Slater determinant of the $A$-body system
\begin{eqnarray}
\Phi_{0} = \frac{1}{\sqrt{A!}}\ {\det} \left| \phi_{1}(1) \phi_{2}(2) \cdots \phi_{A}(A) \right|,
\label{eq:slater_d}
\end{eqnarray}
where the single particle wave functions $\{ \phi_{\gamma} \}$ are orthonormal.
With help of the linked- and unlinked-term expansion of the Slater determinant in Eq.~(\ref{eq:det_expansion}), the matrix element is expressed as
\begin{eqnarray}
&&{\langle \Phi_{0} | Q | \Phi_{0} \rangle} 
= {\left\langle \Phi_{0} \left| \sum_{i_1 \not= i_2 \not= \cdots \not= i_N =1}^{A} Q_{1}(i_1,\cdots,i_{n_1}) \cdots Q_{\alpha}(i_{n_{\alpha-1}+1},\cdots,i_{N}) \right| \Phi_{0} \right\rangle} 
\nonumber \\
&&\hspace*{10mm}
= {\langle Q_1 \rangle}_{\rm c}\ {\langle \overline{Q_1} \rangle} \nonumber \\
&&\hspace*{10mm}
+ \sum_{2 \le p_2 \le \alpha}\ {\langle Q_1 Q_{p_2} \rangle}_{\rm c}\ {\langle \overline{Q_1 Q_{p_2} } \rangle} \nonumber \\
&&\hspace*{10mm}
+ \sum_{2 \le p_2 < p_3 \le \alpha}\ {\langle Q_1 Q_{p_2} Q_{p_3} \rangle}_{\rm c}\ {\langle \overline{Q_1 Q_{p_2} Q_{p_3} } \rangle} \nonumber \\
&&\hspace*{10mm}
+ \cdots\cdots\cdots\cdots\cdots\cdots \nonumber \\
&&\hspace*{10mm}
+ \sum_{2 \le p_{2} < \cdots < p_{\alpha-1} \le \alpha}\ {\langle Q_1 Q_{p_2} \cdots Q_{p_{\alpha-1}} \rangle}_{\rm c}\ {\langle \overline{Q_1 Q_{p_2} \cdots Q_{p_{\alpha-1}} } \rangle} \nonumber \\
&&\hspace*{10mm}
+ \ {\langle Q_1 Q_{2} \cdots Q_{\alpha-1} Q_{\alpha} \rangle}_{\rm c}.
\label{eq:sum_of_connected_terms_1}
\end{eqnarray}
Here the definitions of above notations are presented as  
\begin{eqnarray}
\begin{split}
{\langle Q_1 \rangle}_{\rm c} 
&= \left\langle \Phi_{0} \left| \sum_{i_1 \not=\cdots \not= i_{n_1} = 1}^{A} Q_1(i_1,\cdots,i_{n_1}) \right| \Phi_{0} \right\rangle_{\rm c},
\\
{\langle \overline{Q_1} \rangle} 
&= \left\langle \Phi_{0} \left| \sum_{i_{n_{1}+1} \not=\cdots \not= i_{N} = 1}^{A} Q_2(i_{n_{1}+1},\cdots,i_{n_{2}}) \cdots Q_{\alpha}(i_{n_{\alpha-1}+1},\cdots, i_{N}) \right| \Phi_{0} \right\rangle,
\\
{\langle Q_1 Q_{2} \rangle}_{\rm c} 
&= {\left\langle \Phi_{0} \left| \sum_{i_1 \not=\cdots \not= i_{n_{2}}= 1}^{A} Q_1(i_1,\cdots,i_{n_1}) Q_2(i_{n_1+1},\cdots,i_{n_2}) \right| \Phi_{0} \right\rangle}_{\scalebox{1.1}{\rm c}}, 
\\
{\langle \overline{Q_1 Q_{2} } \rangle} 
&= {\left\langle \Phi_{0} \left| \sum_{{i_{n_{2}+1} \not=\cdots \not= i_{N}} =1}^{A} Q_3(i_{n_{2}+1},\cdots,i_{n_3}) \cdots Q_{\alpha}(i_{n_{\alpha-1}+1},\cdots,i_{N}) \right| \Phi_{0} \right\rangle}. 
\end{split}
\label{eq:lme}
\end{eqnarray}
The notation of the matrix element with the subscript of `c', ${\langle \prod_{k} Q_{k} \rangle}_{\rm c} = {\langle \Phi_{0} | \sum_{i} \prod_{k} Q_{k} | \Phi_{0}  \rangle}_{\rm c}$, in Eqs.~(\ref{eq:sum_of_connected_terms_1}) and (\ref{eq:lme}) is called the {\it linked matrix element}, in which only the linked integrals in the matrix element ${\langle \Phi_{0} | \sum_{i} \prod_{k} Q_{k} | \Phi_{0}  \rangle}$ are summed up (see also Sec.~\ref{sub:lds}).
Other notations are self-explanatory.

Next we consider the matrix element of the operator product $F_1 F_2 \cdots F_n$ in Eq.~(\ref{eq:F1F2...Fn}) with the Slater determinant $\Phi_{0}$ in Eq.~(\ref{eq:slater_d}).
In Appendix~A, we have proved that $F_1 F_2 \cdots F_n$ is expressed as the sum of the operator product in units of the linked ones (non-separable ones).
On the other hand, the matrix element of the product of non-separable operator with $\Phi_{0}$ is expressed in Eq.~(\ref{eq:sum_of_connected_terms_1}).
With help of Eqs.~(\ref{eq:expansion_F1_Fn}),  (\ref{eq:expansion_F1_Fn_connected_operator}), and (\ref{eq:sum_of_connected_terms_1}), the matrix element of the product operator  $F_1 F_2 \cdots F_n$ with $\Phi_{0}$ is finally presented as 
\begin{eqnarray}
{\langle \Phi_{0} | F_1 F_2 \cdots F_n | \Phi_{0} \rangle}  
&\equiv& {\langle F_1 F_2 \cdots F_n \rangle}  \nonumber \\
&=& {\langle F_1 \rangle}\ {\langle \overline{F_1} \rangle} \nonumber \\
&+& \sum_{2 \le p_2 \le n}\ {\langle F_1 F_{p_2} \rangle}_{\rm c}\ {\langle \overline{F_1 F_{p_2} } \rangle} \nonumber \\
&+& \sum_{2 \le p_2 < p_3 \le n}\ {\langle F_1 F_{p_2} F_{p_3} \rangle}_{\rm c}\ {\langle \overline{F_1 F_{p_2} F_{p_3} } \rangle} \nonumber \\
&+& \cdots\cdots\cdots \nonumber \\
&+& \sum_{2 \le p_{2} < \cdots < p_{n-1} \le n}\ {\langle F_1 F_{p_2} \cdots F_{p_{n-1}} \rangle}_{\rm c}\ {\langle \overline{F_1 F_{p_2} \cdots F_{p_{n-1}} } \rangle} \nonumber \\
&+& {\langle F_1 F_2 \cdots F_n \rangle}_c, 
\label{eq:expansion_F1_Fn_dis_c_recursion}
\end{eqnarray}
where ${\langle O \rangle}_{\rm c} \equiv {\langle \Phi_{0} | O | \Phi_{0} \rangle}_{\rm c}$ represents the linked matrix element for $\langle \Phi_{0} | O | \Phi_{0} \rangle$ (see also Sec.~\ref{sub:lds}), and ${\langle \overline{O} \rangle} \equiv {\langle \Phi_{0} | \overline{O} | \Phi_{0} \rangle}$ stands for the matrix element of the operator $\overline{O}$, where $\overline{O}$ means the operator removed the building block of $O$ from $F_1 F_2 \cdots F_{n}$. 
Equation~(\ref{eq:expansion_F1_Fn_dis_c_recursion})  can be proven by using formulae, Eq.~(\ref{eq:sum_of_connected_terms_1}) etc.
From the recurrence formula in Eq.~(\ref{eq:expansion_F1_Fn_dis_c_recursion}), we get the following formula of the linked matrix element expansion 
\begin{eqnarray}
&&{\langle \Phi_{0} | F_1 F_2 \cdots F_n | \Phi_{0} \rangle}  \equiv\  {\langle F_1 F_2 \cdots F_n \rangle}  \nonumber \\
&&\hspace*{5mm}=\ \sum_{ \{k \} } \ 
\left[ \prod_{1 \le p_1 \le n} { \langle F_{p_1} \rangle }^{k^{(1)}_{p_1}} \right]
\left[ \prod_{1 \le p_1 < p_2 \le n} { {\langle F_{p_1} F_{p_2} \rangle}_{\rm c} }^{k^{(2)}_{p_1 p_2}} \right]
\left[ \prod_{1 \le p_1 < p_2 < p_3 \le n} { {\langle F_{p_1} F_{p_2} F_{p_3} \rangle}_{\rm c} }^{k^{(3)}_{p_1 p_2 p_3}} \right] \nonumber \\
&&\hspace*{10mm}\times\ \cdots\ \times
\left[ \prod_{1 \le p_1 < p_2 \cdots < p_{n-1} \le n} { {\langle F_{p_1} F_{p_2} \cdots  F_{p_{n-1}} \rangle}_{\rm c} }^{k^{(n-1)}_{p_1 p_2 \cdots p_{n-1}}} \right] 
{ {\langle F_{1} F_{2} \cdots  F_{{n}} \rangle}_{\rm c} }^{k^{(n)}_{1 2 \cdots n}},
\label{eq:expansion_F1_Fn_dis_c}
\end{eqnarray}
where $\{ k \} =\{ k^{(1)}_{1}, k^{(1)}_{2}, \cdots, k^{(n)}_{1 2 \cdots n} \}$ 
run under the condition,
\begin{eqnarray}
&&k^{(1)}_{\beta} + {\sum_{p_1 < \beta} k^{(2)}_{p_1 \beta}} + {\sum_{\beta < p_1} k^{(2)}_{\beta p_1} } \nonumber \\
&&\hspace*{6mm}+\ {\sum_{p_1 <p_2 < \beta} k^{(3)}_{p_1 p_2\beta}} + {\sum_{p_1 < \beta <p_2 } k^{(3)}_{p_1 \beta p_2}} + {\sum_{ \beta < p_1 < p_2 } k^{(3)}_{\beta p_1  p_2}} 
\nonumber \\
&&\hspace*{6mm}+\cdots\cdots\cdots \nonumber \\
&&\hspace*{6mm}+\ {\sum_{p_1 < \cdots < p_{m-1} < \beta} k^{(m)}_{p_1 \cdots p_{m-1} \beta}}
+\ {\sum_{p_1 < \cdots < p_{m-2} < \beta < p_{m-1}} k^{(m)}_{p_1 \cdots p_{m-2} \beta p_{m-1}}} \nonumber \\
&&\hspace*{20mm}+\cdots\cdots\ + {\sum_{\beta < p_1 < \cdots < p_{m-1}} k^{(m)}_{\beta p_1 \cdots p_{m-1}}} \nonumber \\
&&\hspace*{6mm}+\cdots\cdots\cdots \nonumber \\
&&\hspace*{6mm}+\ k^{(n)}_{1 2 \cdots n} =1  \ \ \ {\rm with}\ {\beta=1,\cdots,n}.
\label{eq:expansion_F1_Fn_dis_c_conditions}
\end{eqnarray}
In Eq.~(\ref{eq:expansion_F1_Fn_dis_c_recursion}), the matrix element is expanded with reference to the operator $F_1$. 
However, we can expand it with reference to an arbitrary operator $F_{a}$ ($a=1,2,\cdots,n$).

From Eq.~(\ref{eq:expansion_F1_Fn_dis_c_recursion}), we get some useful expressions for an arbitrary symmetric many-body operator $\hat{O}$ except $\hat{O}\not= \hat{1}$,
\begin{eqnarray}
&&{\langle F^{n} \rangle} = \sum_{k=0}^{n-1}
\frac{(n-1)!}{k! (n-k-1)!}\ {\langle F^{n-k} \rangle}_{\rm c}\ {\langle F^{k} \rangle},\nonumber\\
&&{\langle \hat{O} F^{n-1} \rangle} = \sum_{k=1}^{n}
\frac{(n-1)!}{(k-1)! (n-k)!}\ {\langle \hat{O} F^{k-1} \rangle}_{\rm c}\ {\langle F^{n-k} \rangle}, \nonumber\\
&&{\langle  F^{n-1} \hat{O} \rangle} = \sum_{k=1}^{n}
\frac{(n-1)!}{(k-1)! (n-k)!}\ {\langle F^{k-1} \hat{O} \rangle}_{\rm c}\ {\langle F^{n-k} \rangle}, \nonumber\\
&&{\langle F^{n_1} \hat{O} F^{n_2} \rangle} 
= \sum_{k=0}^{n_1+n_2}\ \sum_{\substack{{k_1,k_2}\\k_1+k_2=k}} 
\frac{n_{1}!}{k_{1}! (n_{1} -k_{1})!}\ \frac{n_{2}!}{k_{2}! (n_{2} -k_{2})!}
{\langle F^{k_1} \hat{O} F^{k_2} \rangle_{\rm c}} {\langle F^{n_1+n_2-k} \rangle},
\nonumber 
\end{eqnarray}
where ${\langle F^{0} \rangle} = 1$.
Some of them are presented in Sec.~\ref{sub:bepn}.

\end{document}